# COMMUNITY DETECTION ANALYSIS OF SPATIAL TRANSCRIPTOMICS DATA

Charles Zhao[1]

[1]Department of Statistics & Operations Research, University of North Carolina - Chapel Hill

## Abstract

The spatial transcriptomics (ST) data produced by recent biotechnologies, such as CosMx and Xenium, contain huge amount of information about cancer tissue samples, which has great potential for cancer research via detection of *community*: a collection of cells with distinct cell-type composition and similar neighboring patterns. But existing clustering methods do not work well for community detection of CosMx ST data, and the commonly used *k*NN compositional data method shows lack of informative neighboring cell patterns for huge CosMx data. In this article, we propose a novel and more informative *disk compositional data* (DCD) method, which identifies neighboring patterns of each cell via taking into account of ST data features from recent new technologies. After initial processing ST data into DCD matrix, a new innovative and interpretable *DCD-TMHC* community detection method is proposed here. Extensive simulation studies and CosMx breast cancer data analysis clearly show that our proposed DCD-TMHC method is superior to other methods. Based on the communities detected by DCD-TMHC method for CosMx breast cancer data, the logistic regression analysis results demonstrate that DCD-TMHC method is clearly interpretable and superior, especially in terms of assessment for different stages of cancer. These suggest that our proposed novel, innovative, informative and interpretable DCD-TMHC method here will be helpful and have impact to future cancer research based on ST data, which can improve cancer diagnosis and monitor cancer treatment progress.

**March 25, 2025**

[1]*Corresponding author*; E-mail: charlesz@unc.edu

## 1. Introduction

The recent development of *spatial transcriptomics* (ST) technologies, such as CosMx machine and Xenium machine released in December 2022, has shown great potential for improvement of cancer diagnosis and cancer treatment; see He *et al.* (2022) and review article by Jin *et al.* (2024). CosMx machine, i.e., *CosMx Spatial Molecular Imager*, produces data consisting of gene expression at the single-cell level with detailed spatial location information, as well as images of tissues and cells (He *et al.* 2022). The earlier generation of, or something similar to, ST data, such as CODEX (Goltsev *et al.* 2018), Visium (Stahl *et al.* 2016), GeoMx (started in 2019, see Smith *et al.* 2024), etc., does not offer the same data information as CosMx. For instance, CODEX focuses on protein marker, which is considered as part of gene expression, thus gives data different from ST data by CosMx, though having the same form (Williams *et al.* 2022; Chen *et al.* 2023). Visium and GeoMx data provide average information of many cells, not at single-cell level; see review papers by Chen *et al.* (2023) and Cilento *et al.* (2024).

In cancer research, studies have shown that data with the same form as ST data, i.e., data with cell type and spatial location, can help identify groups of cell types with special characteristics that are associated with survival rate and different treatments. In colorectal cancer study, Schurch *et al.* (2020) used CODEX data to identify 9 distinct cellular neighborhoods with different characteristics of the immune tumor microenvironment, which led to the discovery of the enrichment of PD-$1^+$CD$4^+$T cells in a specific granulocyte cellular neighborhood that was associated with better survival for high-risk patients. In triple negative breast cancer study, Shiao *et al.* (2024) also used CODEX data to identify 12 distinct spatial districts which helped detect and analyze different immune response to treatment therapy for two groups of patients. In this article, we define ***community*** as a collection of cells with distinct cell-type composition and similar neighboring patterns. This means that each cell in the community has similar neigh-boring cell types as other cells in the same community, which characterizes the tumor microenvironment. Note that a community contains multiple types of cells, and may be located at different parts of the data, not just concentrated in one spatial area.

In such context, obviously *community detection* for ST data obtained by new biotechnologies, such as CosMx and Xenium, is of great importance in cancer research. Considering the information of neighboring cell types, discussion given in Section 2 shows that the initial step should process ST data into *compositional data* (Quinn *et al.* 2018), then use appropriate clustering method for community detection. The clustering method used by Shiao *et al.* (2024) was Leiden method (Traag *et al.* 2019) based on vectors formed by graph neural network, but Leiden method is not applicable to compositional data. The clustering method used by Schurch *et al.* (2020) was $k$-means method (Hartigan and Wong, 1979; Lloyd, 1982; Marron and Dryden, 2021) for compositional data.

Other existing clustering methods which are applicable to compositional data are: *Elbow* $k$-means method (Schubert, 2023), *Gap* $k$-means method (Tibshirani *et al.* 2001), Mclust algorithm (Fraley and Raftery, 2002; Scrucca *et al.* 2023), DBSCAN algorithm (Hahsler *et al.* 2019), and HDBSCAN algorithm (Campello *et al.* 2013; Campello *et al.* 2015). However, ST data produced by CosMx is usually very huge, and some of these existing clustering methods cannot handle it, while some are under too many parametric and not justifiable assumptions.

Due to these reasons, this article first proposes a novel compositional data method, called *disk compositional data* (DCD) which particularly considers the features of ST data produced by recent new technologies, then proposes an innovative community detection method, called *DCD-TMHC method*, where TMHC is constructed based on 2-means method and *hierarchical clustering method* (Ward, 1963; Murtagh and Contreras, 2012). The simulation studies in Section 3 show that our proposed DCD-TMHC method consistently performs superior to above listed clustering methods. Note that this DCD-TMHC method is generally applicable to ST data or the data with the same form as ST data.

Further, we apply the proposed DCD-TMHC method and other three applicable methods to analyze a CosMx breast cancer dataset, which was originally produced by the CosMx machine at *The UNC Lineberger Comprehensive Cancer Center* and correctively processed by *NanoString* company. Based on the data analysis results, we compare the performance of these methods via logistic regression model. It is obvious that our proposed DCD-TMHC method is clearly superior to other methods, especially in terms of assessment for different stages of cancer. Thus, the DCD-TMHC method proposed in this article will be helpful and have impact to future cancer research based on ST data for improvement of cancer diagnosis and cancer treatment.

The rest of this article is organized as follows. Section 2 reviews ST data and existing methods, and proposes novel community detection method DCD-TMHC; Section 3 presents simulation study results; Section 4 conducts CosMx breast cancer data analysis using proposed DCD-TMHC method and 3 alternative methods, then makes comparison via logistic regression model; and Section 5 gives discussion and conclusion.

## 2. Community Detection of ST Data

*Spatial Transcriptomics* (ST) techniques have existed over almost the past decade, but only were available at the institutions where they were developed. Recently, new techniques, such as CosMx and Xenium, have been commercialized and made ST technology more accessible (Williams *et al.* 2022). ST quantifies messenger RNA transcripts for gene expression at the single-cell level with spatial context. In cancer research, it is often very difficult to extract DNA from formalin-fixed paraffin-embedded tissues, because they are obtained previously from a patient's treatment, thus not "fresh" samples; see review paper by Cilento *et al.* (2024). Now, ST technologies by CosMx and Xenium can handle and process such tissues. CosMx allows quantification of 1000+ RNA and 64+ protein targets, and Xenium can be run on a panel with maximum 480 gene markers.

In particular, the above mentioned CosMx breast cancer data provide 19-25 *fields of view* (FOVs) from each breast cancer tissue sample, which are small selected rectangular regions, and the location (*x, y*) of cell *C* in all FOVs from the same tissue sample is determined based on one origin. Within each FOV, there are thousands of cells and their spatial information, and the small grid portions of cells have good quality with highly detailed information. Below, Figure 1 shows an example of 25 FOVs produced by CosMx from one patient's primary breast cancer sample. Notice that some FOVs are very close to each other, while others are more distant from each other.

**Figure 1. Example of FOVs from Cancer Sample 1: AER8-TTP1**

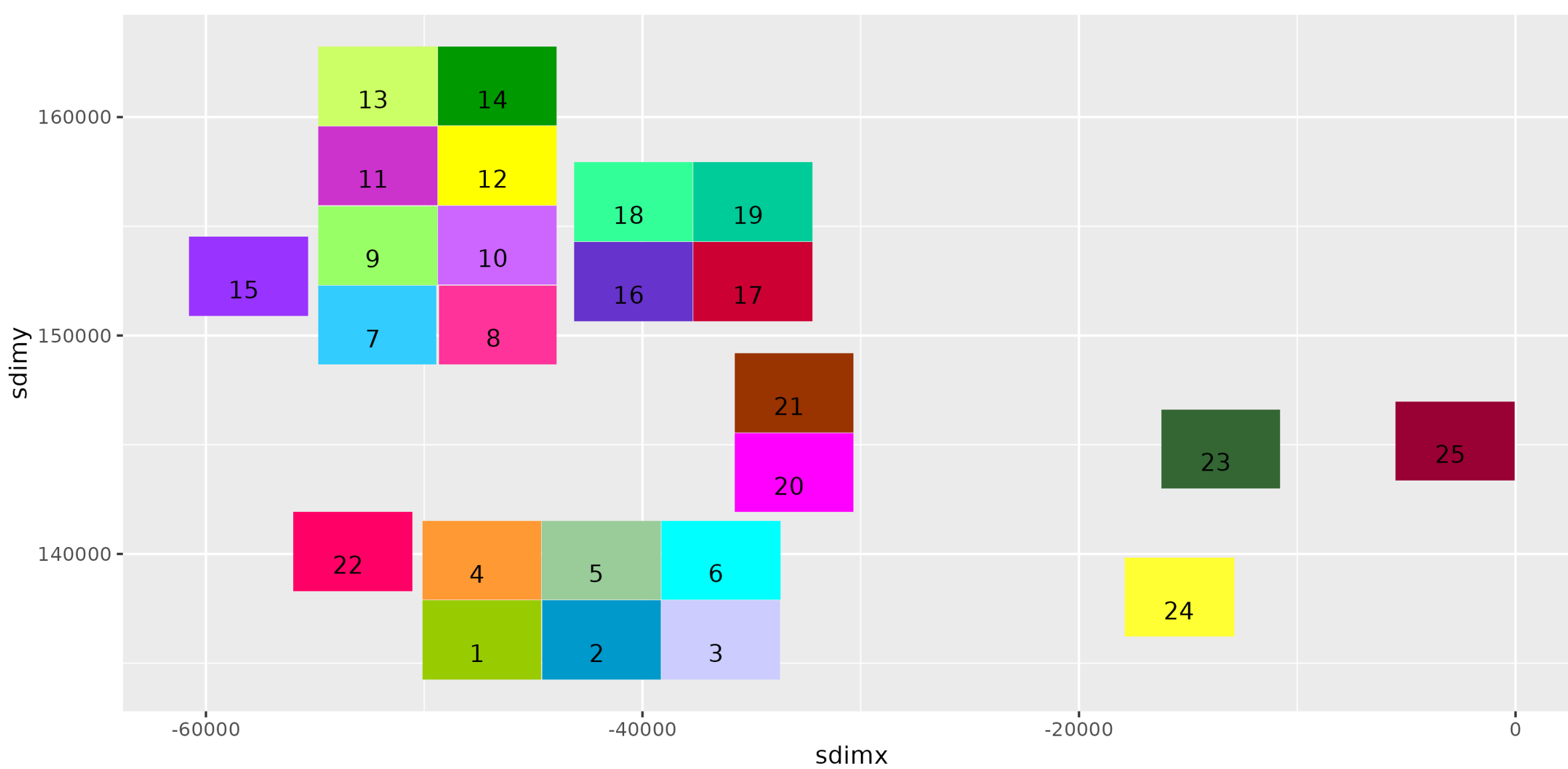

As mentioned in Section 1, **community detection** for ST data obtained by new biotechnologies is of great importance in cancer research. To identify neighboring pattern of each cell, the initial step of community detection needs to conduct data processing to turn ST data into *compositional data* (Quinn *et al.* 2018). In Schurch *et al.* (2020) for colorectal cancer and in Enfield *et al.* (2024) for lung cancer, both had the data form with cell type and spatial location like ST data, and both had their data processed into compositional data first, then used *k*-means method (Hartigan and Wong, 1979; Lloyd, 1982; Marron and Dryden, 2021) to obtain communities for their studies.

### 2.1 Disk Compositional Data Matrix

In this subsection, we review *kNN compositional data matrix*, then propose our *disk compositional data matrix* and make comparison.

#### *k*NN Compositional Data

Both Schurch *et al.* (2020) and Enfield *et al.* (2024) used *k-nearest neighbor* (*k*NN) method to obtain their compositional data by using $k = 10$, which is the most commonly used method in current computational biology literature. We describe the *kNN compositional data matrix* obtained from one sample as follows.

Suppose that sample $S_1$ contains $N_1$ cells with a total of $m$ different cell types. For cell $C_i$ in sample $S_1$, let $k_{ij}$ be the total number of cell type $j$ among $k$ nearest cells in sample $S_1$ around $C_i$, then the *k*NN compositional data matrix for sample $S_1$ is given by:

$$\boldsymbol{C}_1 = (\boldsymbol{v}_1, \dots, \boldsymbol{v}_{N_1})^T \tag{1}$$

where $\boldsymbol{C}_1$ is an $N_1 \times m$ matrix, and

$$\boldsymbol{v}_i = \left(\frac{k_{i1}}{k}, \cdots, \frac{k_{im}}{k}\right)^\top, \quad i = 1, \cdots, N_1. \tag{2}$$

If ST data $\boldsymbol{D}$ contains $q$ samples $S_1, \cdots, S_q$ with total numbers of cells $N_1, \cdots, N_q$, respectively, and each sample $S_j$ contains the same $m$ different cell types, then the *k*NN compositional data matrix for $\boldsymbol{D}$ is given by:

$$\boldsymbol{C} = (\boldsymbol{C}_1^T, \cdots, \boldsymbol{C}_q^T)^T \tag{3}$$

where $\boldsymbol{C}$ is an $N \times m$ matrix with $N = N_1 + \cdots + N_q$, and each $N_j \times m$ matrix $\boldsymbol{C}_j$ is obtained in the same way as shown in above equation (1). Note that matrix $\boldsymbol{C}$ reflects the neighboring cell patterns and restricts overly large values by equation (2).

In Schurch *et al.* (2020), $q = q_s = 35$ cancer samples with a total cell number $N = 132,437$ give sample cell average as $3,784$. In Enfield *et al.* (2024), $q = q_s = 198$ cancer samples with a total cell number $N = 2.3$ million give sample cell average as $11,616$. In comparison, the above mentioned CosMx breast cancer data contains $q_s = 8$ cancer tissue samples with a total cell number $N = 601,634$, which gives a much bigger sample cell average as $75,204$ and a total of $q = 192$ FOVs. The *k*NN compositional data restricts the cell number in the neighbor of any cell $C$ to be always the chosen constant $k$, which can cause the loss of information in the region with very high cell density, and can also cause misleading information for always counting the $k$ nearest cells. The distance between cell $C$ and the $k$th nearest cell can be very tiny, and can also be very huge. These motivate the construction of a novel, more flexible, more adaptive and more informative approach for processing ST data into compositional data as follows.

**Disk Compositional Data**

Choosing a suitable radius $r$, for any cell $C_i$ in sample $S_1$ with distance to the boundary of $S_1$ (if the boundary exists obviously) no less than $r/2$, consider the following disk:

$$B_r(C_i) = \{C \mid || C - C_i || \leq r\} \tag{4}$$

which contains all cells in sample $S_1$ centered around cell $C_i$ within radius $r$, and let $n_{ij}$ be the total number of cell type $j$ included in disk $B_r(C_i)$, then the *disk compositional data matrix* for sample $S_1$ is given by $\tilde{N}_1 \times m$ matrix: $D_1 = \left(u_1, \cdots, u_{\tilde{N}_1}\right)^T$, where

$$\boldsymbol{u_i} = \left(\frac{n_{i1}}{n_i}, \cdots, \frac{n_{im}}{n_i}\right)^\top, \quad i = 1, \cdots, \tilde{N}_1 \tag{5}$$

with $n_i = \sum_{j=1}^{m} n_{ij}$ as the total cell number in disk $B_r(C_i)$ and $\tilde{N}_1$ as the total cell number in sample $S_1$ with distance to the boundary no less than $r/2$. For ST data $\boldsymbol{D}$ containing $q$ samples $S_1, \cdots, S_q$ and each sample $S_j$ containing the same $m$ different cell types, the disk compositional data matrix for $\boldsymbol{D}$ is given by:

$$\boldsymbol{\mathcal{C}}_D = \left(\boldsymbol{\mathcal{D}}_1^\top, \cdots, \boldsymbol{\mathcal{D}}_q^\top\right)^\top \tag{6}$$

where $\boldsymbol{C}_D$ is an $N_D \times m$ matrix with $N_D = \tilde{N}_1 + \cdots + \tilde{N}_q$, and each $\tilde{N}_j \times m$ matrix $\boldsymbol{D}_j$ is obtained in the same way as above $\boldsymbol{D}_1$.

The exclusion of certain cells too close to the sample boundary, such as FOVs in CosMx data, in above process of obtaining disk compositional data matrix $\boldsymbol{C}_D$ is because their corresponding disks given by (4) often contain too few cells. But in cases without obvious boundary like FOVs, such as simulation studies in Section 3 of this article, no cells in data $\boldsymbol{D}$ are excluded to obtain matrix $\boldsymbol{C}_D$ given by (6).

**Comparison of *k*NN and Disk Compositional Data**

In Figure 2 (a) below, one FOV from above mentioned CosMx breast cancer data is displayed, where the use of different radius $r = 125, 250, 500$, respectively, is indicated. Note that near the boundary of the FOV, there are very few cells, and that for smaller radius, disk $B_r(C_i)$ contains too few cells at times, while for bigger radius, disk $B_r(C_i)$ contains too many cells. Thus, it is important to choose a suitable radius.

In Figure 2 (b), for above mentioned CosMx breast cancer data with $N = 601, 634$ cells, we use $r = 250$ to obtain the proposed disk compositional data matrix $\boldsymbol{C}_D$ with $N_D = 524, 366$, then we display the histogram based on all total cell number $n_i$'s in disk $B_r(C_i)$'s for all $q = 192$ FOVs.

In Figure 2 (c), for above mentioned CosMx breast cancer data with $N = 601, 634$ cells, since to our best knowledge $k = 10$ is the most commonly used, we use $k = 10$ to obtain the $k$NN compositional data matrix $\boldsymbol{C}$, then we display the histogram based on all distance $d_{i(10)}$ between cell $C_i$ and its 10th nearest cell for all $q = 192$ FOVs.

Comparing histograms in Figure 2, we see that from Figure 2 (b), overwhelming disk $B_r(C_i)$'s have total cell number $n_i$'s in the range of $20 - 60$, but from Figure 2 (c), we see that if using radius $r = d_{i(10)}$ for cell $C_i$, more than 400,000 disk $B_r(C_i)$'s with $r \approx 150$ contain only 10 cells. Moreover, Figure 2 (c), with red line indicating the maximum value of $d_{i(10)}$'s, shows that even using radius $r = d_{i(10)} > 1,700$, some disks $B_r(C_i)$'s only contain 10 cells. Thus, it is obvious that disk compositional data matrix is far more informative and accurate than $k$NN compositional data matrix for huge ST data.

**Figure 2. FOV Example and Histograms of Disk and *k*NN Methods**

**(a) 1st FOV from Cancer Sample 1: AER8-TTP1**

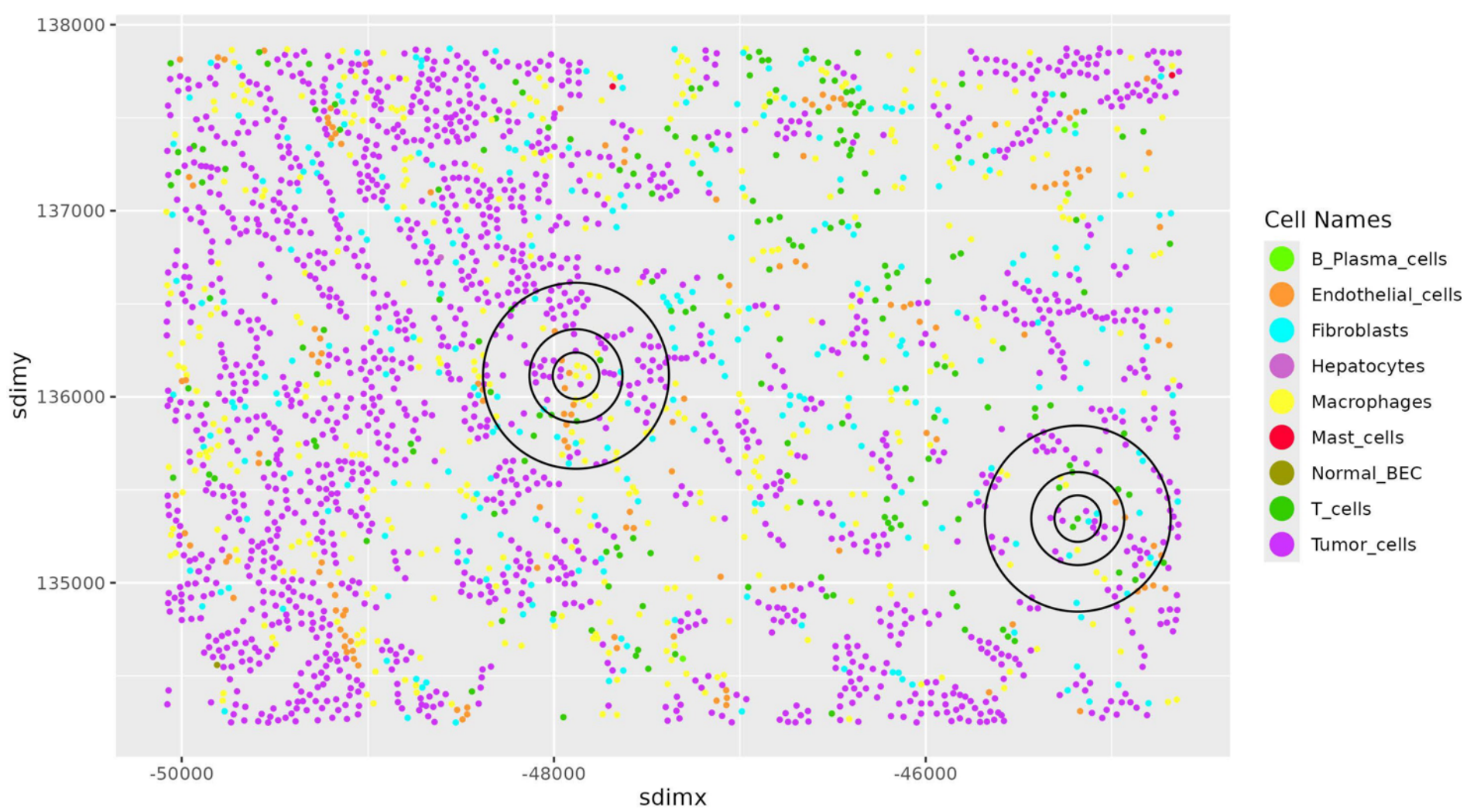

**(b) Histogram of Cell Number in Disk**

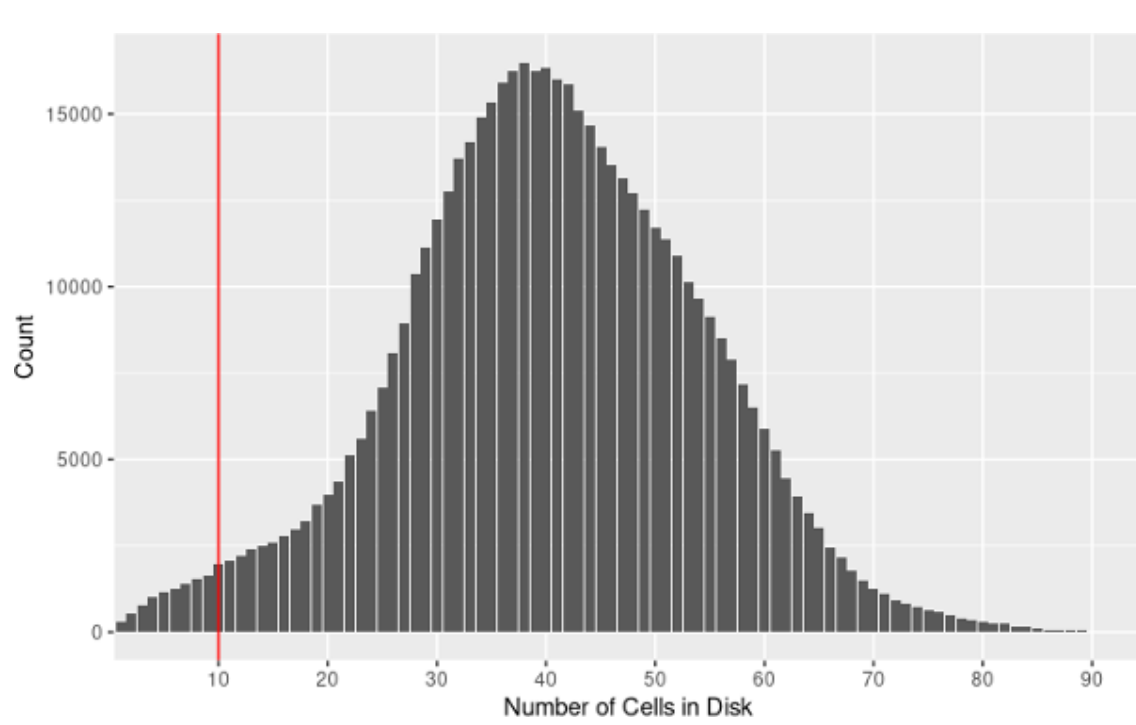

**(c) Histogram of 10th Nearest Distance**

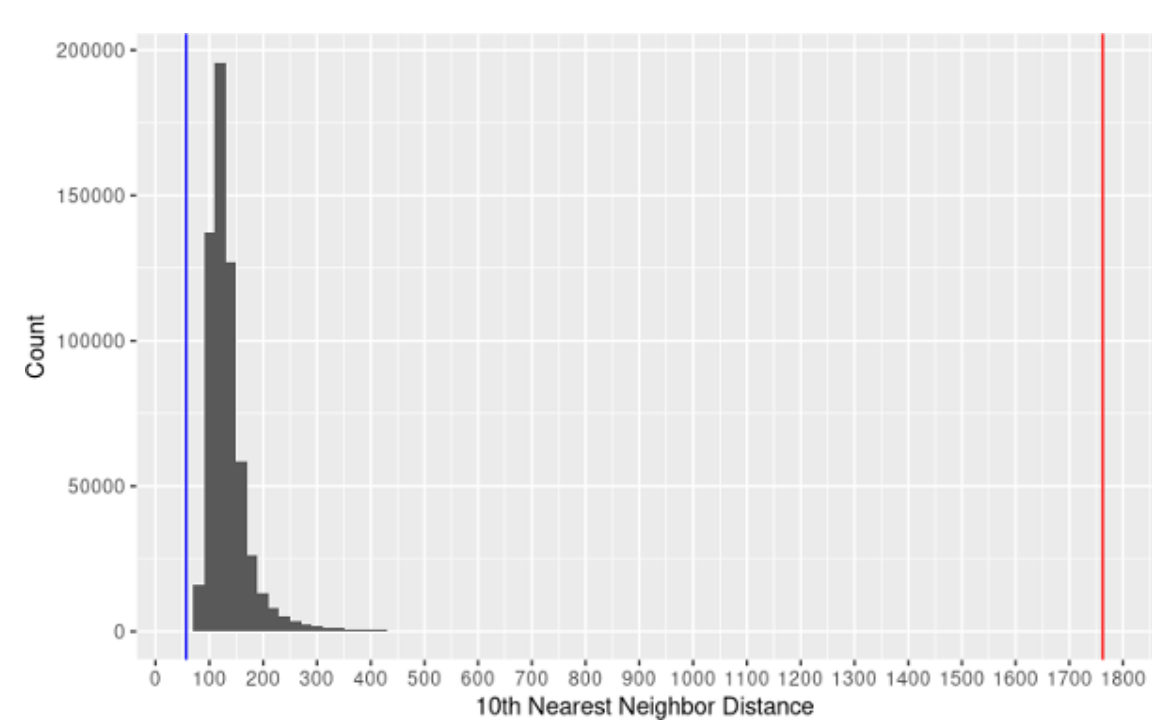

## 2.2 Existing Clustering Methods

After processing ST data into compositional data matrix, the next step is to use appropriate clustering method for community detection. This subsection gives a brief review of several existing and commonly used clustering methods in literature.

***Hierarchical Clustering Method:*** Suppose that $G_1, \cdots, G_n, G_{n+1}$ is a partition of dataset $\boldsymbol{D}$, we obtain a further partition $\hat{G}_1, \cdots, \hat{G}_n$ of $\boldsymbol{D}$ via the following equation:

$$\{\hat{G}_1, \cdots, \hat{G}_n\} = \underset{E_1,\cdots,E_n}{\arg\min} \sum_{i=1}^{n} \sum_{\boldsymbol{v}_j \in E_i} \|\boldsymbol{v}_j - \bar{\boldsymbol{v}}_i\|^2 \tag{7}$$

where $E_1, \cdots, E_n$ is any partition of $\boldsymbol{D}$ with one of them as the union of two $G_i$'s and the rest remaining the same, and $\bar{\boldsymbol{v}}_i$ is the average of all points in $E_i$. If $N$ is the total points in $\boldsymbol{D}$, for a selected $k$, the *hierarchical clustering* (HC) method (Ward, 1963; Murtagh and Contreras, 2012; Marron and Dryden, 2021) is to obtain $k$ clusters by starting with a partition of $\boldsymbol{D}$ with $N$ clusters, repeatedly using (7) and stopping at $n = k$. This method is super time-consuming, thus cannot handle the huge ST data.

***k-Means Method:*** For a selected $k$, the *k-means method* (Hartigan and Wong, 1979; Lloyd, 1982; Marron and Dryden, 2021) is to obtain $k$ clusters $G_1, \cdots, G_k$ of dataset $\boldsymbol{D}$ by solving the following equation:

$$\{G_1, \cdots, G_k\} = \underset{E_1,\cdots,E_k}{\arg\min} \sum_{i=1}^{n} \sum_{\boldsymbol{v}_j \in E_i} \|\boldsymbol{v}_j - \bar{\boldsymbol{v}}_i\|^2 \tag{8}$$

where $E_1, \cdots, E_k$ is any partition of $\boldsymbol{D}$ and $\bar{\boldsymbol{v}}_i$ is the average of all points in $E_i$.

***Elbow k-Means Method:*** The *Elbow method* is using graphical method to choose optimal $k$ for the $k$-means method given by above equation (8); see Schubert (2023).

***Gap k-Means Method:*** The *Gap statistics method* is using statistical method to choose optimal $k$ for the $k$-means method via equation (8); see Tibshirani *et al.* (2001).

***Mclust Algorithm:*** The *Mclust algorithm* is to fit the data by 14 different Gaussian mixture models, then use the *Bayesian Information Criterion* (BIC) to identify the optimal number of clusters by choosing the best Gaussian mixture model; see Fraley and Raftery (2002), Scrucca *et al.* (2023).

***DBSCAN Algorithm:*** The *density-based spatial clustering of applications with noise* (DBSCAN) algorithm uses density-based method to identify clusters, in which process, two parameters *eps* and *minPts* are involved; see Hahsler *et al.* (2019).

***HDBSCAN Algorithm:*** The *hierarchical density-based spatial clustering of applications with noise* (HDBSCAN) algorithm uses density-based method and minimum spanning tree to identify clusters, in which process, one parameter *minPts* is involved; see Campello *et al.* (2013), Campello *et al.* (2015).

Both DBSCAN and HDBSCAN algorithms are super time-consuming. Other existing methods include *Seurat method* (Hao *et al.* 2021), *UMAP-Mclust method* (He *et al.* 2022), etc., which are incompatible for compositional data, and our simulation studies and data analysis show that these methods do not work well for large ST data.

### 2.3 Data Transformation and SigClust

Before proposing our novel and innovative DCD-TMHC method for community detection of ST data in this article, we first need to describe *Aitchison log-ratio transformation* and *SigClust method*, which are used in the process of our newly developed method.

***Aitchison Log-Ratio Transformation:*** The *Aitchison log-ratio transformation* transfers the compositional data from a simplex back to real Euclidean space, which in many situations gives better results for clustering; see (Aitchison, 1982; 1986). However, it has also been pointed out that in certain situation such transformation may not be necessary; see Quinn *et al.* (2018).

***SigClust:*** *Statistical significance of clustering* (SigClust) method is hypothesis test:

$$H_0 : \text{Data from a single Gaussian distribution} \quad vs. \quad H_1 : H_0 \text{ does not hold} \tag{9}$$

which can be used to determine whether any two clusters are significantly different or if we should put them together to be one cluster; see Liu *et al.* (2008), Huang *et al.* (2015).

### 2.4 Proposed DCD-TMHC Method

**Step 1:** Process ST data into disk compositional data matrix, then transfer it using Aitchison log-ratio transformation;

**Step 2:** Use 2-means method consecutively to obtain clusters of not too big size as shown in Figure 3, where SigClust is used to make sure that the splitted clusters are distinct, and any two non-distinct clusters are considered as one community;

**Step 3:** For distinct clusters obtained in Step 2, apply HC method shown in Figure 4, where SigClust is used on each large enough node of the dendrogram to determine whether the split should be kept: (a) if not or not large enough, stop at the node and treat it as one community; (b) if yes, continue SigClust down to the next node.

**Remark 1.** In practice or simulations studies, the "not too big" in Step 2 is determined by computation power, and "not large enough" in Step 3 uses a chosen number $K_1$ to stop at Step 3 (a). The studies in this article all used super computer at University of North Carolina at Chapel Hill. Moreover, above Step 2 and related Figure 3 suggest an alternative clustering method, called *successive 2-means* (STM) method, which stops at the node if "not large enough" for chosen number $K_1$ or if one of its two splitted clusters has size less than a chosen number $K_2$.

**Figure 3. Successive 2-Means Method**

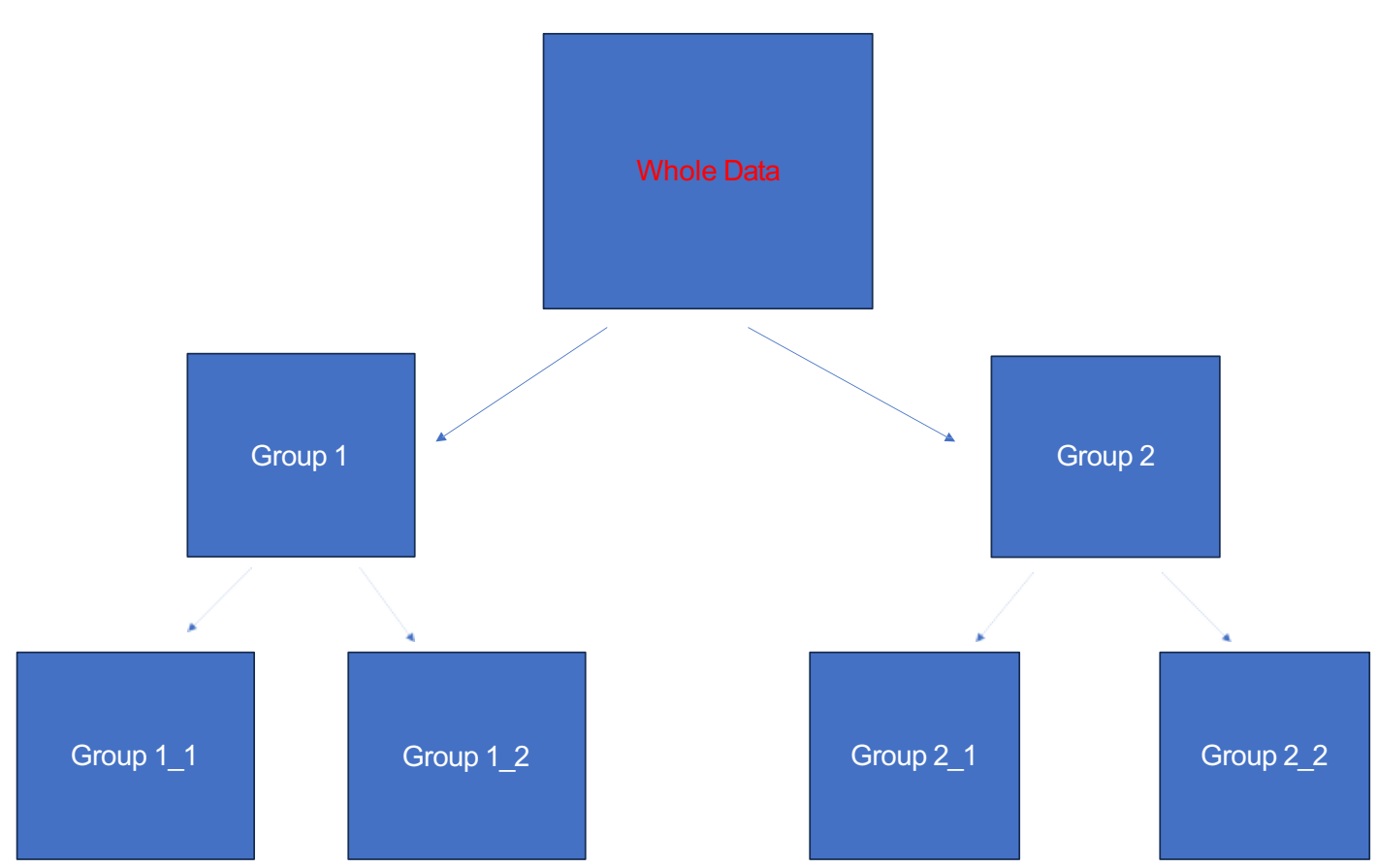

**Figure 4. Step 3 of DCD-TMHC Method**

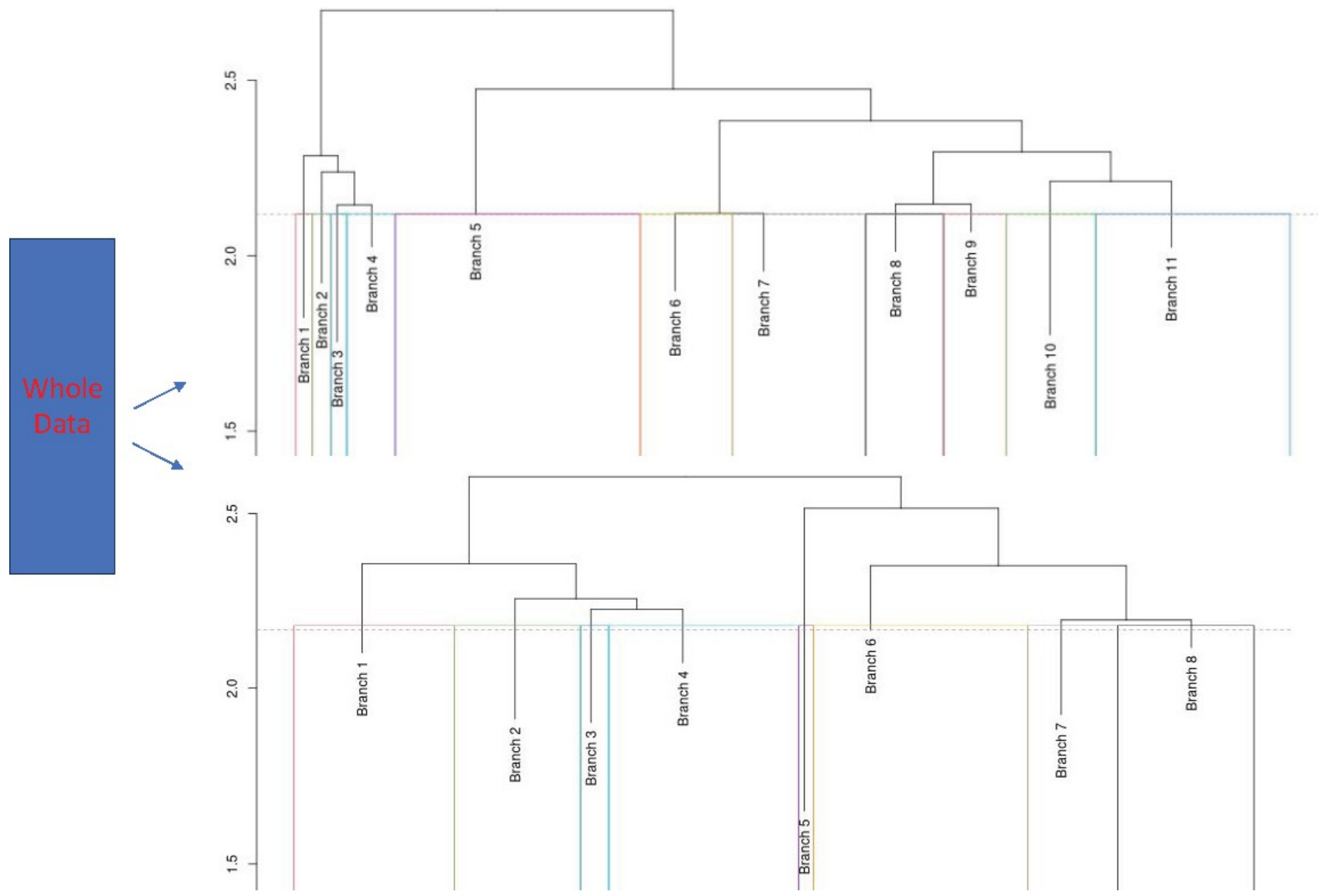

## 3. Simulation Studies

This section presents some simulation results on community detection using various existing methods listed in Section 2 and our proposed DCD-TMHC method, plus the alternative STM method. There are 5 different simulation study settings which are summarized in Table 1. Figure 5 displays the visualization of these 5 simulation settings, and Table 2 (a)-(e) present the summary of simulation results.

In Table 1, we denote $U\{m_1, m_2\}$ as the discrete uniform distribution between positive integers $m_1$ and $m_2$, $U(a, b)$ as the continuous uniform distribution with support interval $(a, b)$, and $N(\mu, \sigma^2)$ as the normal distribution with mean $\mu$ and variance $\sigma^2$, where for simplicity, unit 1 = 1000 is used for parameters of distribution notations. For each simulation setting, the number of cells is generated using discrete uniform distribution listed on the top, except different discrete uniform distributions are used in Community 4 of Simulation 1 and in Community 7 of Simulation 5, respectively. For instance, notation "Cell 1 (12,462)" in Community 1 of Simulation 1 means a total number of 12,462 Cell

Type 1 is generated from $U\{10, 25\}$, where 10 and 25 represent 10,000 and 25,000, respectively. Also, under Simulation 2, notations "Cell 2 (50%=7,370)" in Community 2 and "Cell 2 (50%=7,404)" in Community 3 mean that 50% of the number generated from $U\{10, 15\}$ for Cell Type 2 is used in Communities 2 and 3, respectively.

**Table 1. Five Simulation Study Settings**

| Intended Community | **Simulation 1**<br>$U\{10, 25\}$ | **Simulation 2**<br>$U\{10, 15\}$ | **Simulation 3**<br>$U\{100, 150\}$ | **Simulation 4**<br>$U\{90, 120\}$ | **Simulation 5**<br>$U\{150, 200\}$ |
|---|---|---|---|---|---|
| 1 | Cell 1 (12,462):<br>$X, Y \sim N(0, 250)$ | Cell 1 (11,016):<br>$X, Y \sim U(-.25, .25)$ | Cell 1 (102,985):<br>$X, Y \sim N(2, 100)$ | Cell 1 (45%=48,330):<br>$X \sim U(-.55, -.35)$<br>$Y \sim U(-.25, .25)$ | Cell 1 (152,985):<br>$X, Y \sim N(2, 100)$ |
| 2 | Cell 2 (12,510):<br>$X, Y \sim N(4, 250)$ | Cell 2 (50%=7,370):<br>$X \sim U(-.25, .25)$<br>$Y \sim U(.5, 1)$ | Cell 2 (98,264):<br>$X, Y \sim N(4, 100) \cap R^c$ | Cell 1 (55%=59,070):<br>Cell 4 (45%=52,538):<br>$X, Y \sim U(-.25, .25)$ | Cell 2 (136,445):<br>$X, Y \sim N(4, 100) \cap R^c$ |
| 3 | Cell 3 (20,418):<br>$X, Y \sim N(10, 250)$ | Cell 2 (50%=7,404):<br>Cell 3 (50%=6,066):<br>$X \sim U(.35, .85)$<br>$Y \sim U(.5, 1)$ | Cell 3 (129,709):<br>$X, Y \sim N(8, 100)$ | Cell 2 (45%=51,474):<br>$X \sim U(-.25, .25)$<br>$Y \sim U(.5, 1)$ | Cell 3 (179,709):<br>$X, Y \sim N(8, 100)$ |
| 4 | $U\{6, 8\}$<br>Cell 1 (6,525):<br>Cell 2 (6,525):<br>$X \sim N(8, 50)$<br>$Y \sim N(0, 50)$ | Cell 3 (50%=6,110):<br>$X \sim U(1.15, 1.65)$<br>$Y \sim U(.5, 1)$ | Cell 2 (31,660):<br>$X, Y \sim N(4, 100) \cap R$<br>Cell 4 (44%=60,512):<br>$X, Y \sim U(4, 4.7)$ | Cell 2 (55%=62,913):<br>Cell 3 (45%=42,648):<br>$X \sim U(.35, .85)$<br>$Y \sim U(.5, 1)$ | Cell 2 (43,479):<br>$X, Y \sim N(4, 100) \cap R$<br>Cell 4 (58%=108,766):<br>$X, Y \sim (4, 4.7)$ |
| 5 | | Cell 4 (50%=5,773):<br>$X \sim U(.7, 1)$<br>$Y \sim U(-.15, .15)$<br>Cell 4 (50%=5,759):<br>$\theta \sim U(0, 2\pi)$<br>$X = .4 \cos \theta + .8$<br>$Y = .25 \sin \theta)$ | Cell 4 (56%=77,016):<br>$X \sim U(5, 6)$<br>$Y \sim U(8.5, 9.2)$ | Cell 3 (55%=52,126):<br>$X \sim U(1.15, 1.65)$<br>$Y \sim U(.5, 1)$ | Cell 4 (42%=78,762):<br>Cell 5 (58%=88,598):<br>$X \sim U(5.5, 6.5)$<br>$Y \sim U(8.5, 9.2)$ |
| 6 | | | | Cell 4 (55%=64,214):<br>$X \sim U(.35, .85)$<br>$Y \sim U(1.1, 1.35)$ | Cell 5 (42%=64,158):<br>$X \sim U(2.5, 4.5)$<br>$Y \sim U(8.5, 9.2)$ |
| 7 | | | | Cell 5 (45%=46,447):<br>$X \sim U(.7, 1)$<br>$Y \sim U(-.15, .15)$<br>Cell 5 (55%=56,770):<br>$\theta \sim U(0, 2\pi)$<br>$X = .4 \cos \theta + .8$<br>$Y = .25 \sin \theta$ | $U\{30, 50\}$<br>Cell 1 (42,635):<br>Cell 2 (39,208)<br>$X \sim N(8, 90)$<br>$Y \sim N(2, 90)$ |
| **Notations** | 1 = 1000 for distribution notations; $R = [3995, 4705] \times [3995, 4705]$ | | | | |

## Figure 5. Dot Plots of 5 Simulation Settings

### (a) Simulation 1

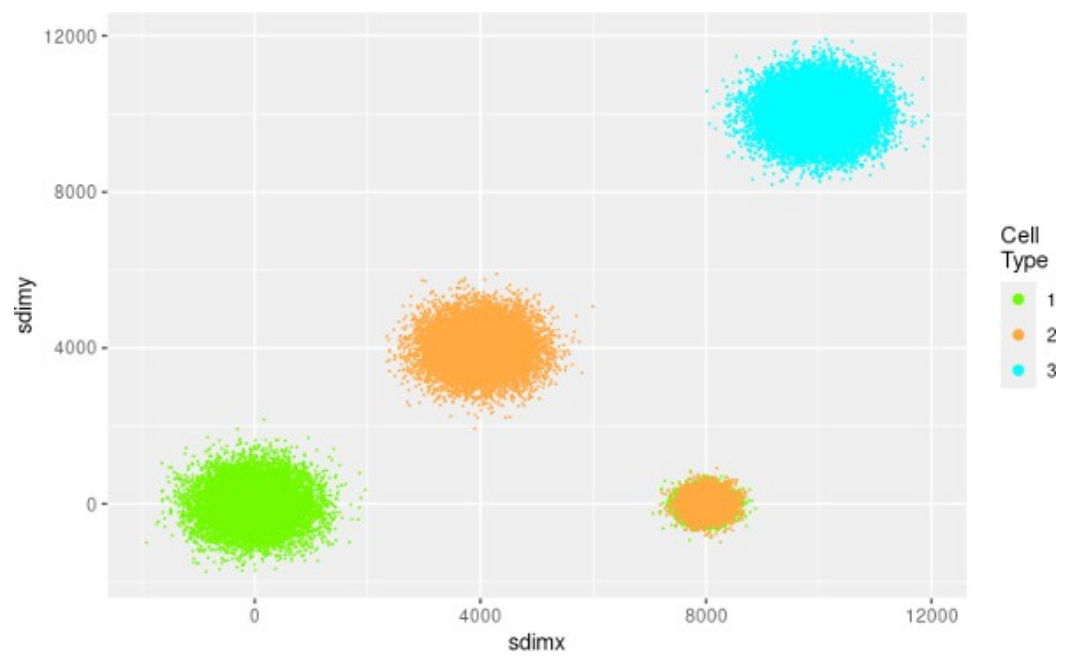

### (b) Simulation 2

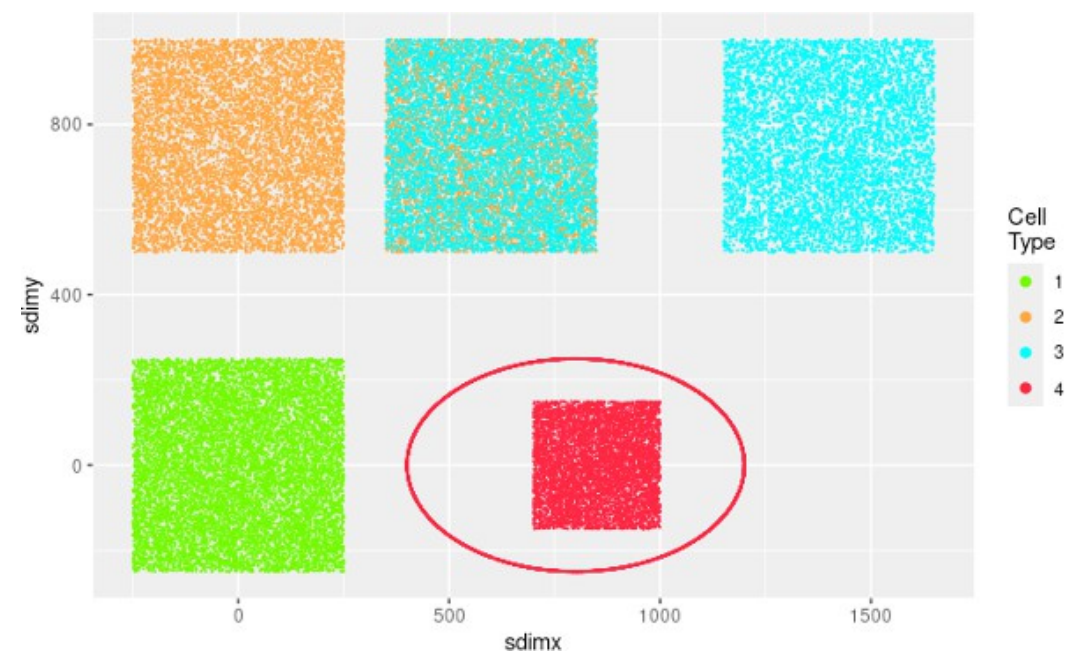

### (c) Simulation 3

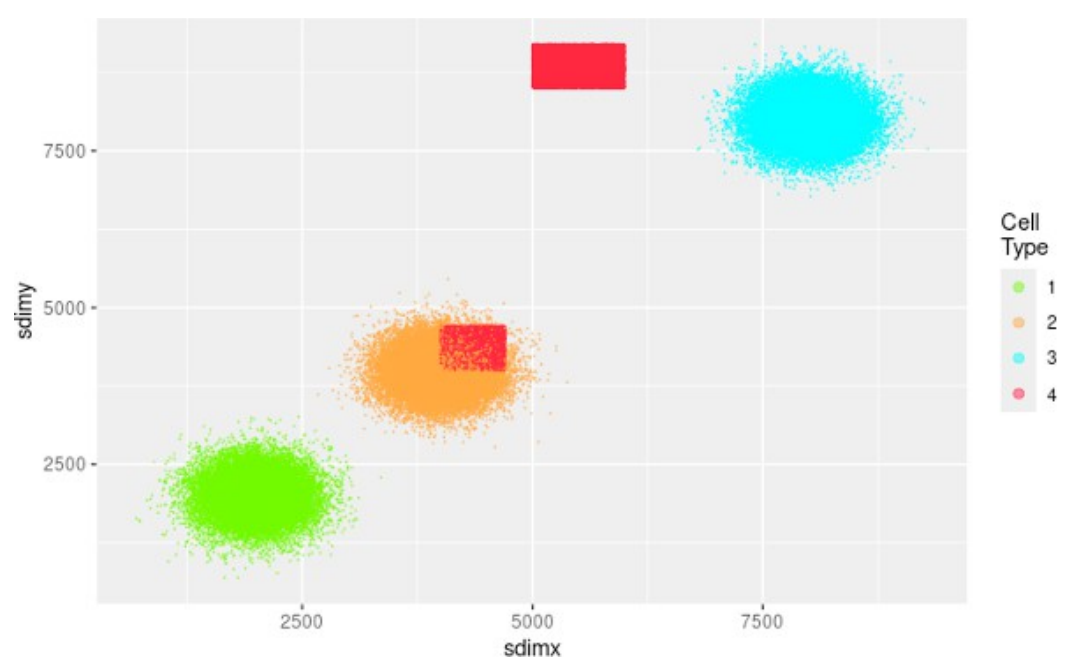

### (d) Simulation 4

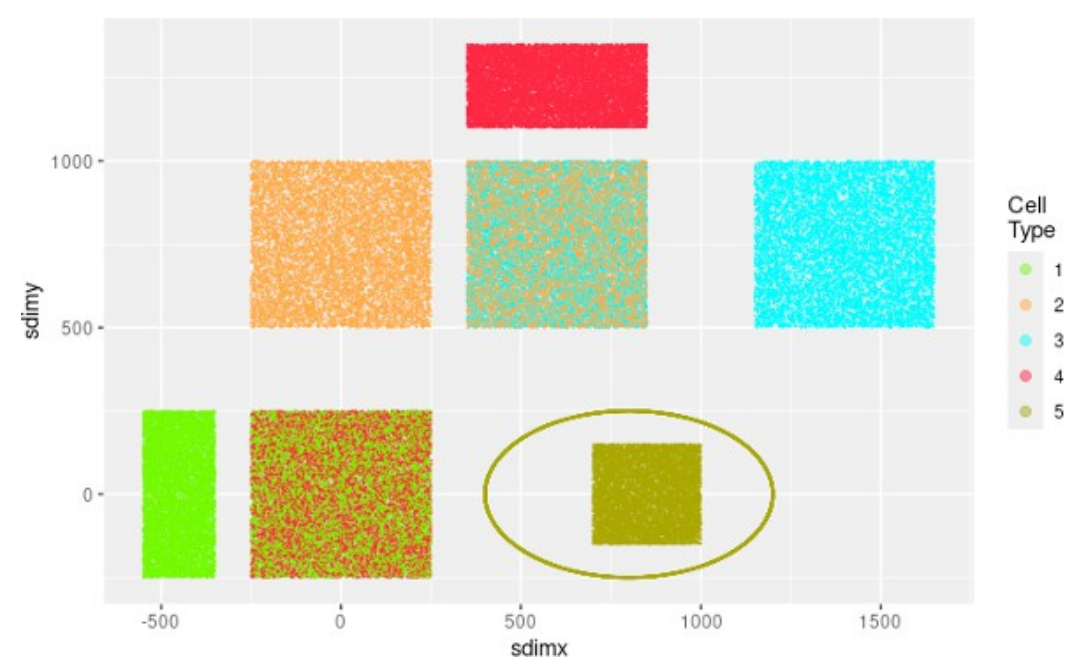

### (e) Simulation 5

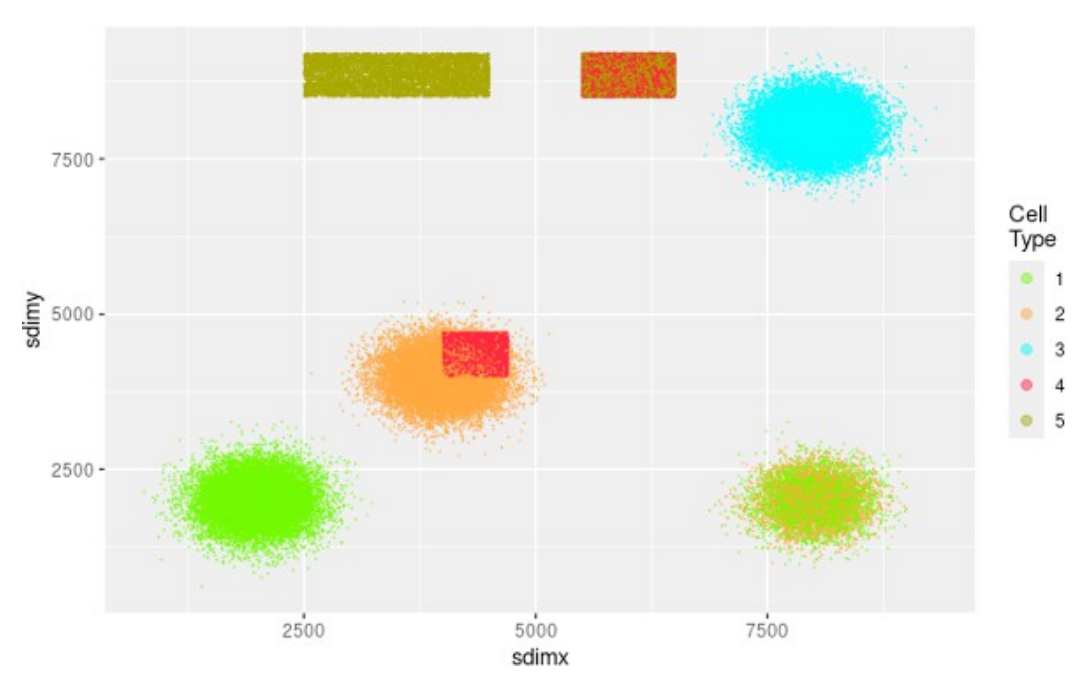

## Table 2 (a). Simulation 1 with $N = 58,440$ Cells

| Method | Detected Community No. | Chosen Parameters | ARI |
|---|---|---|---|
| DCD-TMHC | 4 | $K_1 = 0$ | 0.9815 |
| STM | 5 | $K_1 = 583, K_2 = 150$ | 0.9712 |
| $k$-Means | 4 | | 0.9593 |
| Elbow $k$-Means | 4 | | 0.9593 |
| Gap $k$-Means | 10 | | 0.8783 |
| Mclust | 4 | | 0.9593 |
| HDBSCAN | 2 | $minPts = 1001$ | 0.4590 |
| DBSCAN | 5 | $eps = 0.025, minPts = 1001$ | 0.9337 |

## Table 2 (b). Simulation 2 with $N = 49,498$ Cells

| Method | Detected Community No. | Chosen Parameters | ARI |
|---|---|---|---|
| DCD-TMHC | 5 | $K_1 = 0$ | 0.9997 |
| STM | 5 | $K_1 = 12,000,\ K_2 = 100$ | 0.9104 |
| *k*-Means | 5 | | 0.6178 |
| Elbow *k*-Means | 7 | | 0.8418 |
| Gap *k*-Means | 4 | | 0.7081 |
| Mclust | 5 | | 0.9655 |
| HDBSCAN | 2 | $minPts = 1001$ | 0.2647 |
| DBSCAN | 9 | $eps = 0.02,\ minPts = 1001$ | 0.8348 |

## Table 2 (c). Simulation 3 with $N = 500,146$ Cells

| Method | Detected Community No. | Chosen Parameters | ARI |
|---|---|---|---|
| DCD-TMHC | 7 | $K_1 = 60,000$ | 0.9206 |
| STM | 6 | $K_1 = 60,000,\ K_2 = 10,000$ | 0.8930 |
| *k*-Means | 5 | | 0.6267 |
| Elbow *k*-Means | 9 | | 0.6234 |
| Gap *k*-Means | NA | | NA |
| Mclust | 5 | | 0.8791 |
| HDBSCAN | NA | | NA |
| DBSCAN | 16 | $eps = 0.01,\ minPts = 1001$ | 0.8870 |

## Table 2 (d). Simulation 4 with $N = 536,530$ Cells

| Method | Detected Community No. | Chosen Parameters | ARI |
|---|---|---|---|
| DCD-TMHC | 7 | $K_1 = 0$ | 0.9998 |
| STM | 9 | $K_1 = 60,000,\ K_2 = 10,000$ | 0.7982 |
| *k*-Means | 7 | | 0.9758 |
| Elbow *k*-Means | 7 | | 0.9758 |
| Gap *k*-Means | NA | | NA |
| Mclust | 7 | | 0.9765 |
| HDBSCAN | NA | | NA |
| DBSCAN | 9 | $eps = 0.02,\ minPts = 2001$ | 0.9727 |

## Table 2 (e). Simulation 5 with $N = 934,745$ Cells

| Method | Detected Community No. | Chosen Parameters | ARI |
|---|---|---|---|
| DCD-TMHC | 10 | $K_1 = 80,000$ | 0.8337 |
| STM | 9 | $K_1 = 125,000,\ K_2 = 5000$ | 0.7954 |
| *k*-Means | 7 | | 0.8902 |
| Elbow *k*-Means | 7 | | 0.8902 |
| Gap *k*-Means | NA | | NA |
| Mclust | 7 | | 0.8894 |
| HDBSCAN | NA | | NA |
| DBSCAN | 9 | $eps = 0.015,\ minPts = 2001$ | 0.7782 |

For 5 simulation settings given in Table 1, we first process the data into disk compositional data matrix given by equation (6), but Aitchison data transformation is not used here because the data contain too many zeros. Table 2 (a)-(e) only include simulation results for some existing clustering methods, our proposed DCD-TMHC method and alternative method STM, because HC method is too time-consuming, and methods, such as Seurat, UMAP-Mclust, etc., are incompatible for compositional data. The chosen parameters listed in Table 2 (a)-(e) seem to be the best based on our various experimental testings during the simulation studies.

Note that Table 1 gives the *intended* community number $k$ for each simulation setting, thus the same number $k$ is used for $k$-means and Mclust methods, while all other methods determine the number of communities on their own, and "NA" indicates no results available for certain methods. Also, note that Figure 5 displays the visualization of 5 simulation settings, which agrees the actual community numbers with Simulations 1, 2 and 4. For Simulation 3, the intended community number is 5, but Figure 5 (c) clearly shows that the actual community number is 7-8 because the region with overlapping colors of orange and red has some part more dense than other part, and the boundary between colors orange and red has some part with solid orange surrounding red, while other part with sparse orange surrounding red. Similarly, Simulation 5 has 7 as intended community number, but Figure 5 (e) suggests the actual community number is 9-10.

To check the accuracy of community detection, a good measure is the *adjusted rand index* (ARI) (Hubert and Arabie 1985), which ranges in interval $(-1, 1)$ and indicates high accuracy with number close to 1. In Table 2 (a)-(e), the ARIs are computed for all listed methods assuming the intended communities listed in Table 1 as the truth.

**Remark 2.** In Table 2 (a)-(e), our proposed DCD-TMHC method is obviously superior to any other methods, because it consistently and correctly detects the actual community numbers, and it has the highest ARI for Simulations 1, 2 and 4 which give the accurate communities. From the simulation results in Table 2 (a)-(e), it is easy to see that Gap $k$-means and HDBSCAN methods do not work well. Moreover, Mclust method does not work well either, because without given community number $k$, this method has 1-9 as default community numbers and the finally determined community number is based on the highest BIC out of 14 different Gaussian mixture models; see Figure 6 for BIC plot under Simulation 1 setting, which gives one as the detected community number.

**Figure 6. BIC Plot of Mclust in Simulation 1**

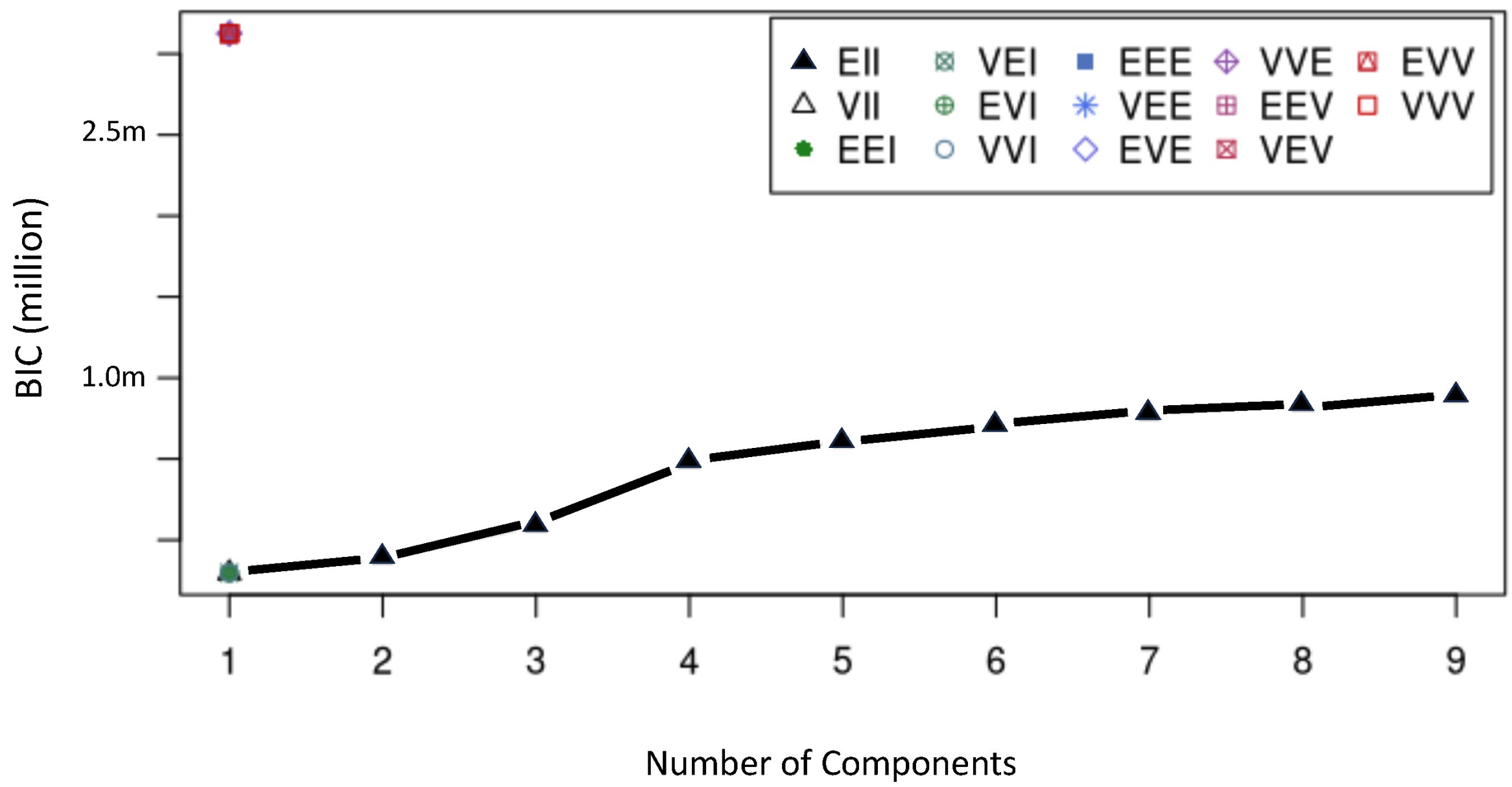

## 4. CosMx Data Analysis

This section applies our proposed DCD-TMHC method, along with a few other methods, to analyze a CosMx breast cancer dataset, which was originally produced by the CosMx machine at *The UNC Lineberger Comprehensive Cancer Center* and correctively processed by NanoString company. The goal is to achieve community detection, then analyze its impact to cancer research.

The CosMx data under consideration here consists of eight samples from four people; each of them had one primary breast cancer sample and one metastasis sample at stage IV of cancer. The total cell number from 8 samples is $N$ = 601634, and the total highly cancer research related $m = 9$ cell types under consideration are: B-plasma, endothelial, fibroblasts, hepatocytes, macrophages, mast, normal-BEC, T and tumor cells.

For each sample, cell types were manually annotated by Leiden method and the expression of cell marker genes, and 19-25 FOVs were manually chosen for the purpose of having many tumor cells, as well as reflecting other cell types present in the data, such as normal breast epithelial (normal-BEC) cells, hepatocytes cells, etc.; see Figure 1 for description of FOVs and see Table 3 for data summary.

**Table 3. CosMx Data Summary**

| Sample Name | Sample No. | Sample Size | Patient No. | Sample Type | Tissue Type |
|---|---|---|---|---|---|
| AER8-TTP1 | 1 | 59,556 | 1 | Primary | Breast |
| AER8-TTM2 | 2 | 57,045 | 1 | Metastasis | Liver |
| AFE4-TTP1 | 3 | 20,495 | 2 | Primary | Breast |
| AFE4-TTM6 | 4 | 84,168 | 2 | Metastasis | Liver |
| RA11-044-PRIM | 5 | 48,092 | 4 | Primary | Breast |
| RA11-044-MET | 6 | 97,895 | 4 | Metastasis | Lung |
| RA11-049-PRIM | 7 | 113,317 | 3 | Primary | Breast |
| RA11-049-MET | 8 | 121,066 | 3 | Metastasis | Liver |

As the initial step of community detection, we use radius $r = 250$ to process the CosMx data $\boldsymbol{D}$ into disk compositional data matrix $\boldsymbol{C}_D$ given by (6) with $N_D = 524, 366$ and $m = 9$, which does not apply the disk given by (4) centered by cell $C_i$'s too close to the boundary of FOV or too isolated. After applying Aitchison log-ratio transformation to matrix $\boldsymbol{C}_D$, we use our proposed DCD-TMHC method, STM method, 10-means method and Elbow $k$-means method to detect communities; see results displayed in Figure 7, which gives the bar charts of $m = 9$ cell types for each detected community.

In Figure 7, Elbow $k$-means method determined $k = 20$, DCD-TMHC method used $K_1 = 1000$, and STM method used $K_1 = 5000$, $K_2 = 100$. In our analysis, we also considered DBSCAN method, but it only detected two communities, thus the result is not included in Figure 7. Other clustering methods, such as Gap $k$-means, HDBSCAN,

Mclust, etc., are not included here due to the discussion given in Remark 2 about simulation results and their poor performance on this CosMx data $\boldsymbol{D}$ in our analysis.

**Figure 7. Bar Charts of Detected Communities**

**(a) DCD-TMHC**

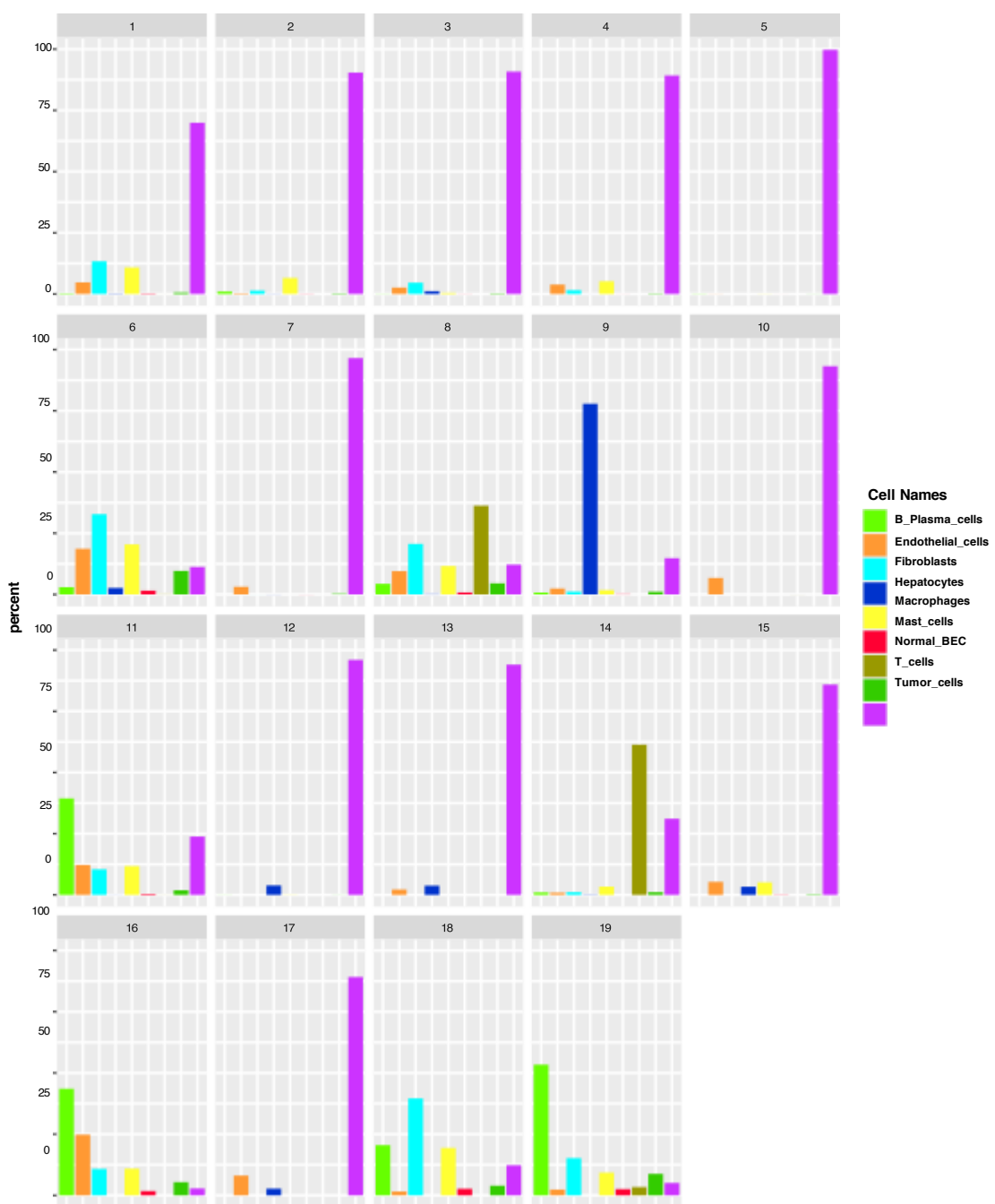

**(b) STM**

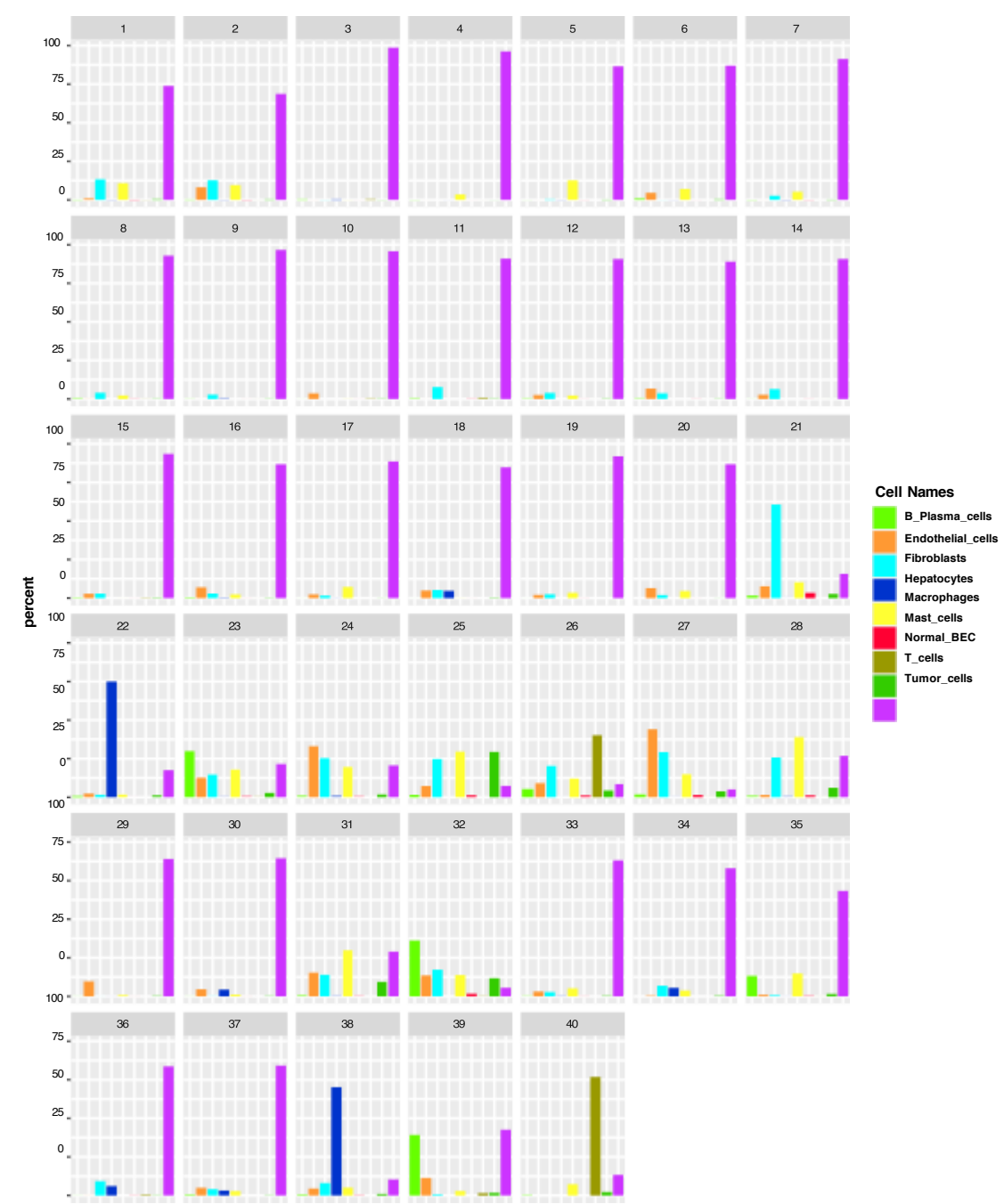

**(c) 10-Means**

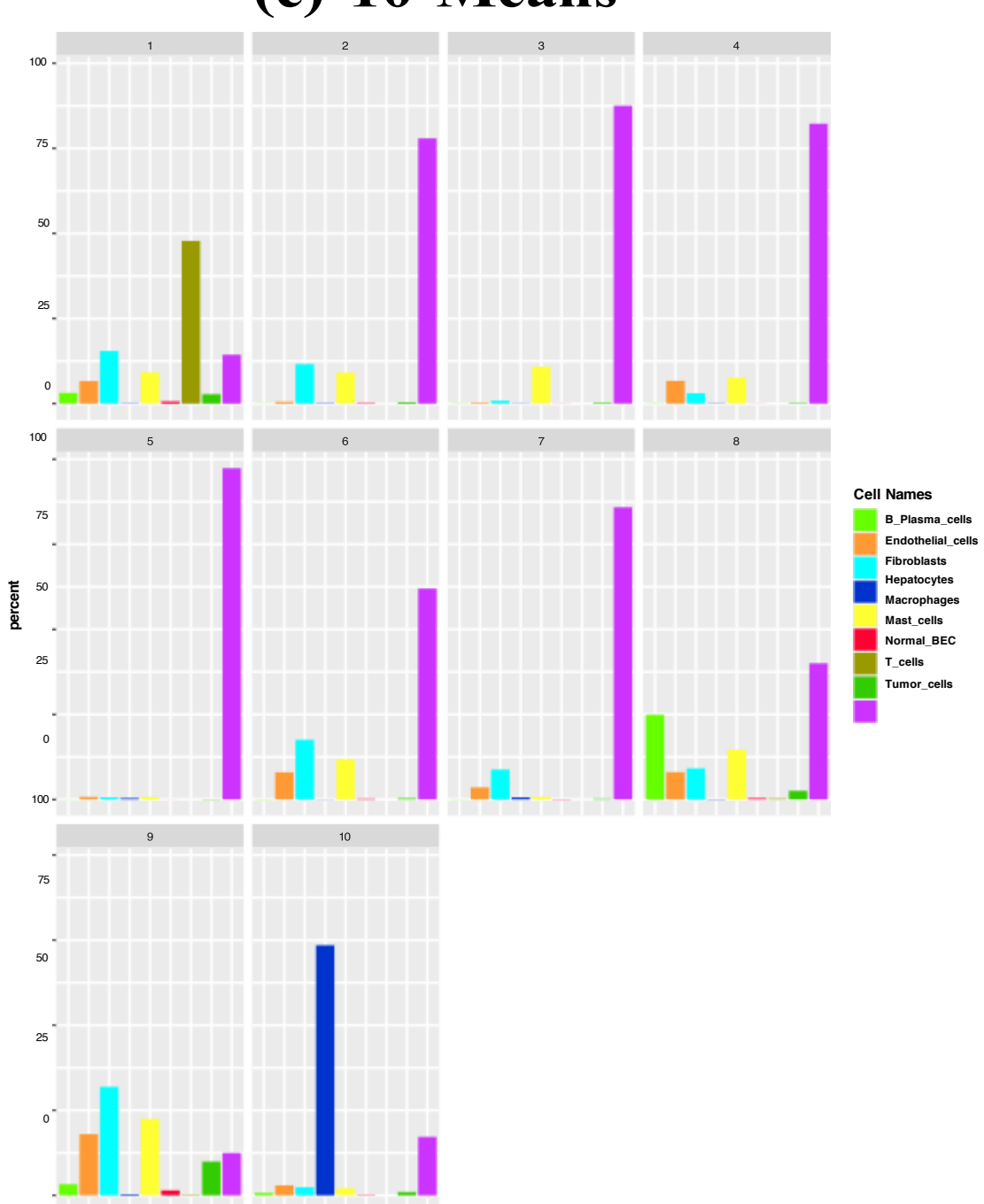

**(d) Elbow 20-Means**

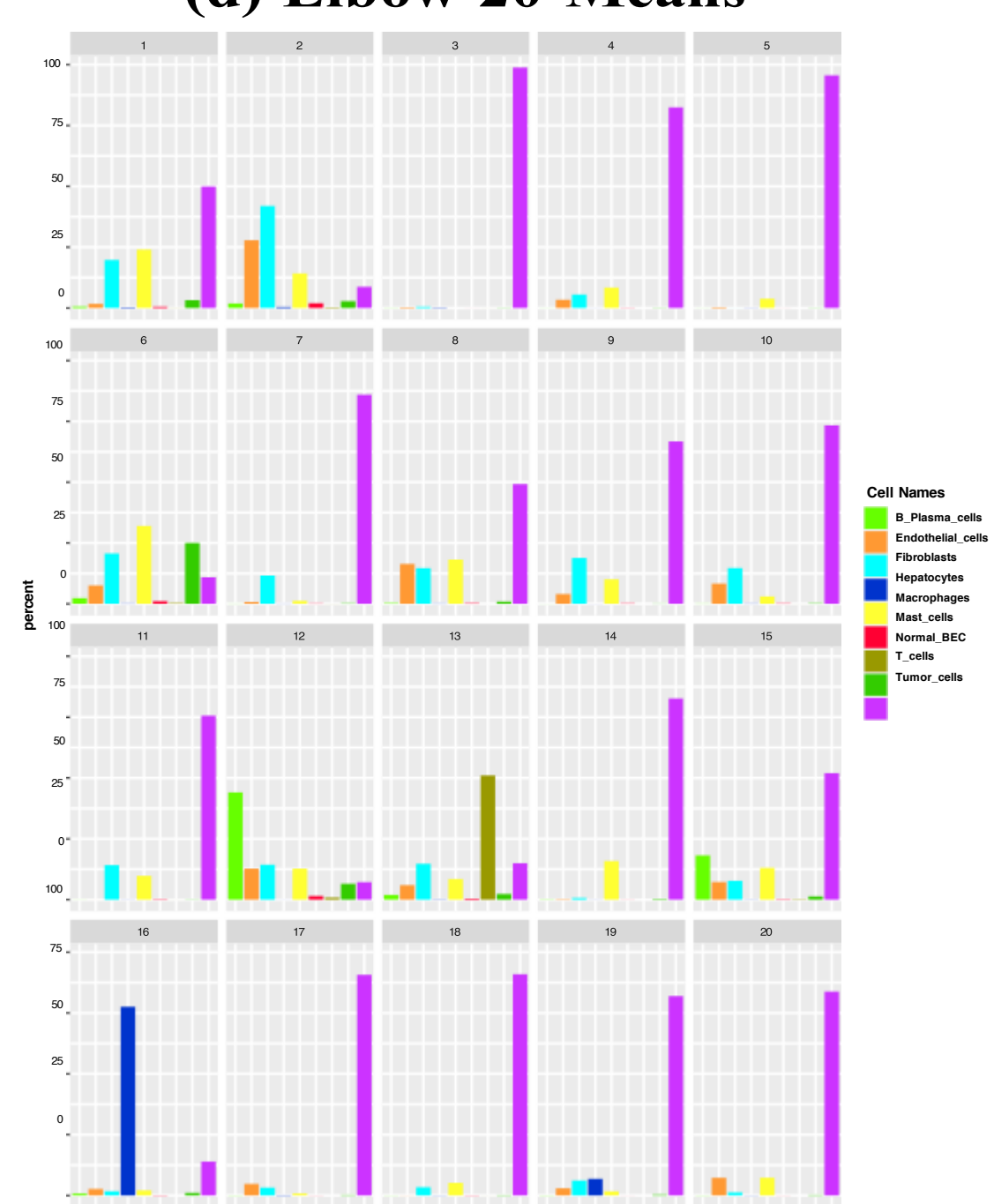

It is interesting to notice that our proposed DCD-TMHC method detected 19 communities for aforementioned CosMx data, while Elbow $k$-means method determined $k = 20$ communities. However, the detected communities among 4 methods displayed in Figure 7 have some similarities and also some noticeable differences. For instance, a bar in each detected community represents the percentage of one cell type within that community. Our DCD-TMHC method detected Community 5 as having the highest percentage of tumor cells, 99.7%, in the community, while STM, 10-means and Elbow $k$-means methods detected communities 3, 5 and 3 as having the highest tumor cell percentages 98.7%, 97.3% and 98.8%, respectively, thus they are quite similar; see Table 4. On the other hand, for immune cell as the sum of B-plasma and T cells, DCD-TMHC method detected Community 19 as having the highest percentage of immune cells, 62.2%, in the community, while STM, 10-means and Elbow $k$-means methods detected communities 32, 8 and 12 as having the highest immune cell percentages 47.95%, 27.4% and 50.69%, respectively, thus they are quite different; see Table 5.

**Table 4. Sample Cell % in Detected Community with Highest Tumor Cell**

| **Method:** DCD-TMHC; **Community (Tumor Cell %):** 5 (99.7%) | | | | | | | | | |
|---|---|---|---|---|---|---|---|---|---|
| Sample No. | 1 | 2 | 3 | 4 | 5 | 6 | 7 | 8 | LR Estimator |
| $x$ | 0.40 | 39.30 | 3.26 | 12.58 | 7.28 | 6.95 | 2.98 | 0.59 | $\hat{\alpha} = 1.375$ |
| $y$ | 1 | 0 | 1 | 0 | 1 | 0 | 1 | 0 | $\widehat{\beta} = -0.219$ |
| **Method:** STM; **Community (Tumor Cell %):** 3 (98.7%) | | | | | | | | | |
| Sample No. | 1 | 2 | 3 | 4 | 5 | 6 | 7 | 8 | LR Estimator |
| $x$ | 0.87 | 42.30 | 3.74 | 14.62 | 7.43 | 6.78 | 2.93 | 0.64 | $\hat{\alpha} = 1.344$ |
| $y$ | 1 | 0 | 1 | 0 | 1 | 0 | 1 | 0 | $\widehat{\beta} = -0.199$ |
| **Method:** 10-Means; **Community (Tumor Cell %):** 5 (97.3%) | | | | | | | | | |
| Sample No. | 1 | 2 | 3 | 4 | 5 | 6 | 7 | 8 | LR Estimator |
| $x$ | 3.06 | 49.58 | 7.28 | 46.69 | 24.86 | 17.54 | 11.85 | 2.69 | $\hat{\alpha} = 1.354$ |
| $y$ | 1 | 0 | 1 | 0 | 1 | 0 | 1 | 0 | $\widehat{\beta} = -0.071$ |
| **Method:** Elbow $k$-Means; **Community (Tumor Cell %):** 3 (98.8%) | | | | | | | | | |
| Sample No. | 1 | 2 | 3 | 4 | 5 | 6 | 7 | 8 | LR Estimator |
| $x$ | 1.17 | 44.69 | 4.66 | 29.73 | 10.86 | 7.87 | 4.82 | 1.36 | $\hat{\alpha} = 1.245$ |
| $y$ | 1 | 0 | 1 | 0 | 1 | 0 | 1 | 0 | $\widehat{\beta} = -0.119$ |

**Table 5. Sample Cell % in Detected Community with Highest Immune Cell**

| **Method:** DCD-TMHC; **Community (Immune B+T Cell %):** 19 (62.2%) | | | | | | | | | |
|---|---|---|---|---|---|---|---|---|---|
| Sample No. | 1 | 2 | 3 | 4 | 5 | 6 | 7 | 8 | LR Estimator |
| $x$ | 0.18 | 0.05 | 0.47 | 0.04 | 0.78 | 0.07 | 0.00 | 0.00 | $\hat{\alpha} = -1.566$ |
| $y$ | 1 | 0 | 1 | 0 | 1 | 0 | 1 | 0 | $\widehat{\beta} = 14.073$ |
| **Method:** STM; **Community (Immune B+T Cell %):** 32 (47.95%) | | | | | | | | | |
| Sample No. | 1 | 2 | 3 | 4 | 5 | 6 | 7 | 8 | LR Estimator |
| $x$ | 1.07 | 0.57 | 5.27 | 0.18 | 1.36 | 0.30 | 0.00 | 0.00 | $\hat{\alpha} = -1.640$ |
| $y$ | 1 | 0 | 1 | 0 | 1 | 0 | 1 | 0 | $\widehat{\beta} = 2.691$ |
| **Method:** 10-Means; **Community (Immune B+T Cell %):** 8 (27.4%) | | | | | | | | | |
| Sample No. | 1 | 2 | 3 | 4 | 5 | 6 | 7 | 8 | LR Estimator |
| $x$ | 0.63 | 0.42 | 4.32 | 0.31 | 3.34 | 10.37 | 0.25 | 0.02 | $\hat{\alpha} = 0.142$ |
| $y$ | 1 | 0 | 1 | 0 | 1 | 0 | 1 | 0 | $\widehat{\beta} = -0.058$ |
| **Method:** Elbow *k*-Means; **Community (Immune B+T Cell %):** 12 (50.6%) | | | | | | | | | |
| Sample No. | 1 | 2 | 3 | 4 | 5 | 6 | 7 | 8 | LR Estimator |
| $x$ | 1.06 | 0.25 | 5.77 | 0.25 | 1.91 | 1.29 | 0.01 | 0.00 | $\hat{\alpha} = -1.163$ |
| $y$ | 1 | 0 | 1 | 0 | 1 | 0 | 1 | 0 | $\widehat{\beta} = 1.251$ |

**Table 6. Sample Cell % in Detected Community with Highest Normal Cell**

| **Method:** DCD-TMHC; **Community (Normal-BEC Cell %):** 14 (61.4%) | | | | | | | | | |
|---|---|---|---|---|---|---|---|---|---|
| Sample No. | 1 | 2 | 3 | 4 | 5 | 6 | 7 | 8 | LR Estimator |
| $x$ | 1.36 | 0.00 | 1.43 | 0.00 | 0.07 | 0.00 | 0.00 | 0.00 | $\hat{\alpha} = -1.386$ |
| $y$ | 1 | 0 | 1 | 0 | 1 | 0 | 1 | 0 | $\widehat{\beta} = 273.171$ |
| **Method:** STM; **Community (Normal-BEC Cell %):** 40 (76.8%) | | | | | | | | | |
| Sample No. | 1 | 2 | 3 | 4 | 5 | 6 | 7 | 8 | LR Estimator |
| $x$ | 0.59 | 0.00 | 0.09 | 0.00 | 0.01 | 0.00 | 0.00 | 0.00 | $\hat{\alpha} = -1.386$ |
| $y$ | 1 | 0 | 1 | 0 | 1 | 0 | 1 | 0 | $\widehat{\beta} = 2660.943$ |
| **Method:** 10-Means; **Community (Normal-BEC Cell %):** 1 (47.8%) | | | | | | | | | |
| Sample No. | 1 | 2 | 3 | 4 | 5 | 6 | 7 | 8 | LR Estimator |
| $x$ | 4.41 | 0.00 | 13.04 | 0.00 | 0.76 | 0.00 | 0.00 | 0.00 | $\hat{\alpha} = -1.386$ |
| $y$ | 1 | 0 | 1 | 0 | 1 | 0 | 1 | 0 | $\widehat{\beta} = 26.804$ |
| **Method:** Elbow *k*-Means; **Community (Normal-BEC Cell %):** 13 (51.0%) | | | | | | | | | |
| Sample No. | 1 | 2 | 3 | 4 | 5 | 6 | 7 | 8 | LR Estimator |
| $x$ | 4.28 | 0.00 | 11.52 | 0.00 | 0.60 | 0.00 | 0.00 | 0.00 | $\hat{\alpha} = -1.386$ |
| $y$ | 1 | 0 | 1 | 0 | 1 | 0 | 1 | 0 | $\widehat{\beta} = 33.942$ |

For many other differences among detected communities by different methods, a particularly noticeable one is that DCD-TMHC method detected two communities, 8 and 14, as having the highest percentage of normal cells, and STM method also detected two such communities: 26 and 40. But, each of 10-means method and Elbow *k*-means method only identifies one such community: Community 1 for 10-means method and Community 13 for Elbow *k*-means method.

As mentioned in Section 1 and Section 2, in Schurch *et al.* (2020) for colorectal cancer and in Enfield *et al.* (2024) for lung cancer, both used 10-means method to detect communities based on their data. Schurch *et al.* (2020) associated certain communities with better survival for high-risk patients, while Enfield *et al.* (2024) identified certain community with high percentage of current smokers which is highly associated with lung cancer. Thus, based on our detected communities in Figure 7, we are interested in their impact and association to cancer research as follows.

In Table 4, for Community 5 detected by our proposed DCD-TMHC method, each of 8 samples listed in Table 3 has the following variables given in Table 4:

$$x_i = \frac{k_i}{N_i} \quad \text{and} \quad y_i = \begin{cases} 1, & \text{if } S_i \text{ is Primary} \\ 0, & \text{if } S_i \text{ is Metastasis} \end{cases}, \qquad i = 1, \cdots, 8 \tag{10}$$

where $k_i$ is the total number of cells in both sample $S_i$ and Community 5 detected by DCD-TMHC method, $N_i$ is the total cell number in sample $S_i$, and the values of $x_i$'s in Table 4 are $x_1$ = 0.40%, $x_2$ = 39.30%, etc. Thus, this is a dataset observed for a binary response variable $Y$ and an explanatory variable $X$, which is naturally analyzed using the following *logistic regression model*: (McCullagh and Nelder, 1989; Agresti, 2002)

$$\pi(x) = P\{Y = 1 \mid X = x\} = \frac{exp(\alpha + \beta x)}{1 + exp(\alpha + \beta x)}, \tag{11}$$

and logistic regression (LR) estimators $\hat{\alpha} = 1.375$ and $\hat{\beta} = -0.219$ are computed based on data (10) and given in Table 4. In turn, the estimated logistic regression curve is:

$$\hat{\pi}(x) = \frac{exp(\hat{\alpha} + \hat{\beta}x)}{1 + exp(\hat{\alpha} + \hat{\beta}x)} \tag{12}$$

which is displayed by a red curve in Figure 8. Note that $\pi(x)$ in equation (11) is the conditional probability of being primary stage of cancer for given value $x$, and above $\hat{\pi}(x)$ is estimated conditional probability of $\pi(x)$.

For the rest of data in Table 4 from three methods of STM, 10-means and Elbow $k$-means, we obtain three estimated logistic regression curves using equations (10)-(12), respectively, and present them in Figure 8 as well for comparison.

**Figure 8. Logistic Regression Curves for Highest Tumor Community**

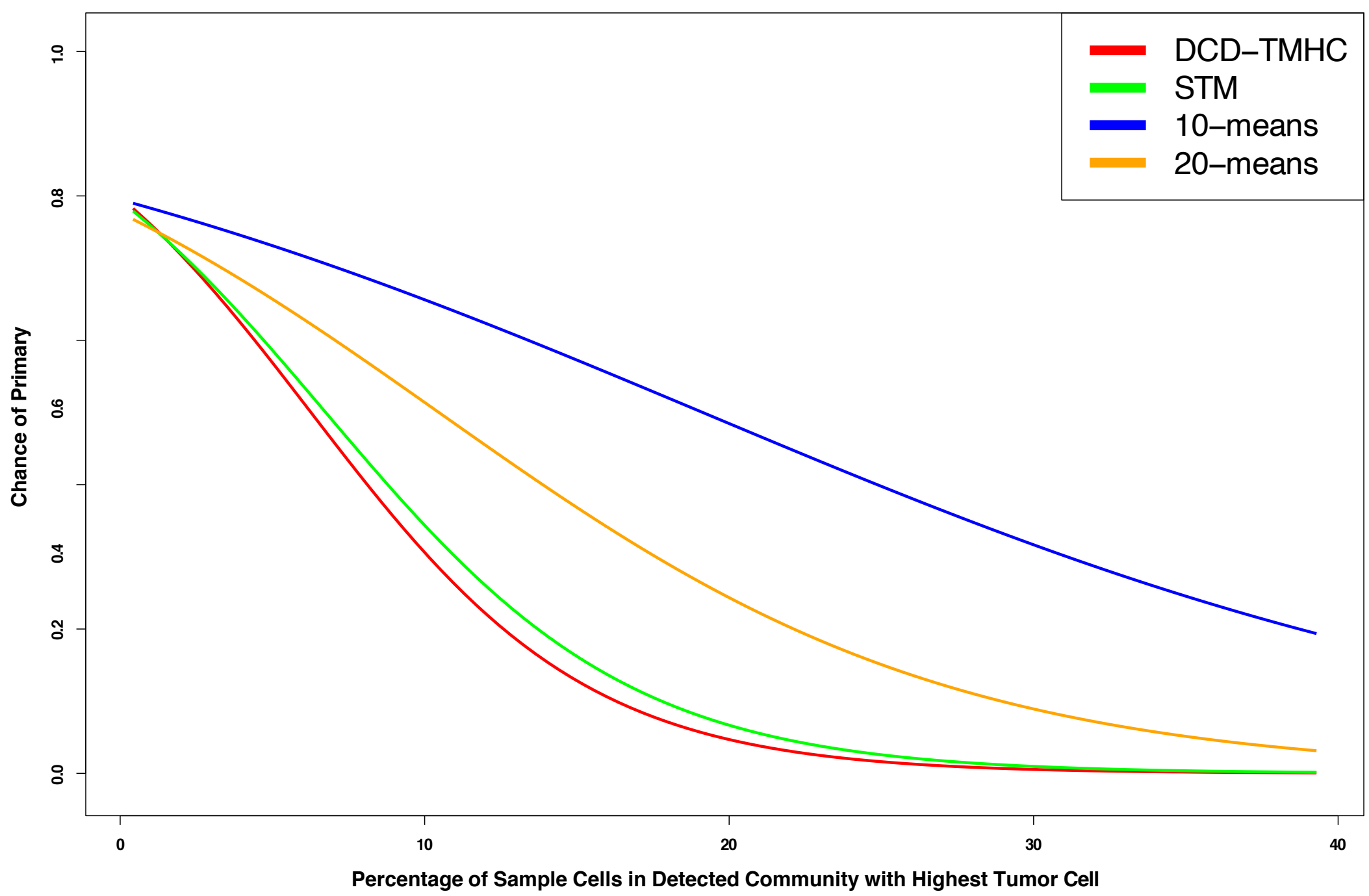

Similarly presented as Table 4, the data in Table 5 and Table 6 are based on detected communities as having the highest percentages of immune cells and normal cells, respectively. Following equations (10)-(12), we obtain estimated logistic regression curves for Table 5 and Table 6, and present them in Figure 9 and Figure 10, respectively.

**Figure 9. Logistic Regression Curves for Highest Immune Community**

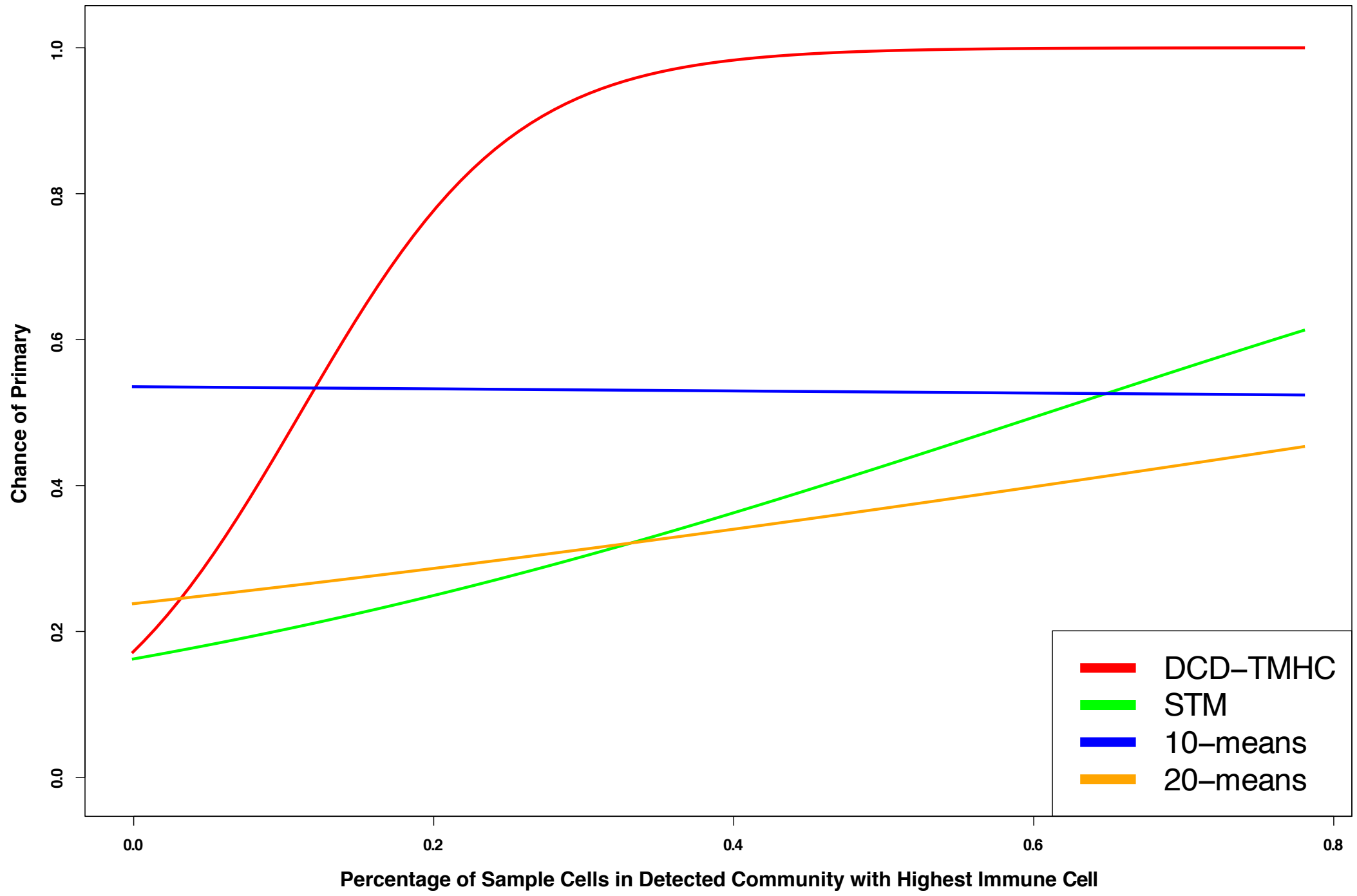

**Figure 10. Logistic Regression Curves for Highest Normal Community**

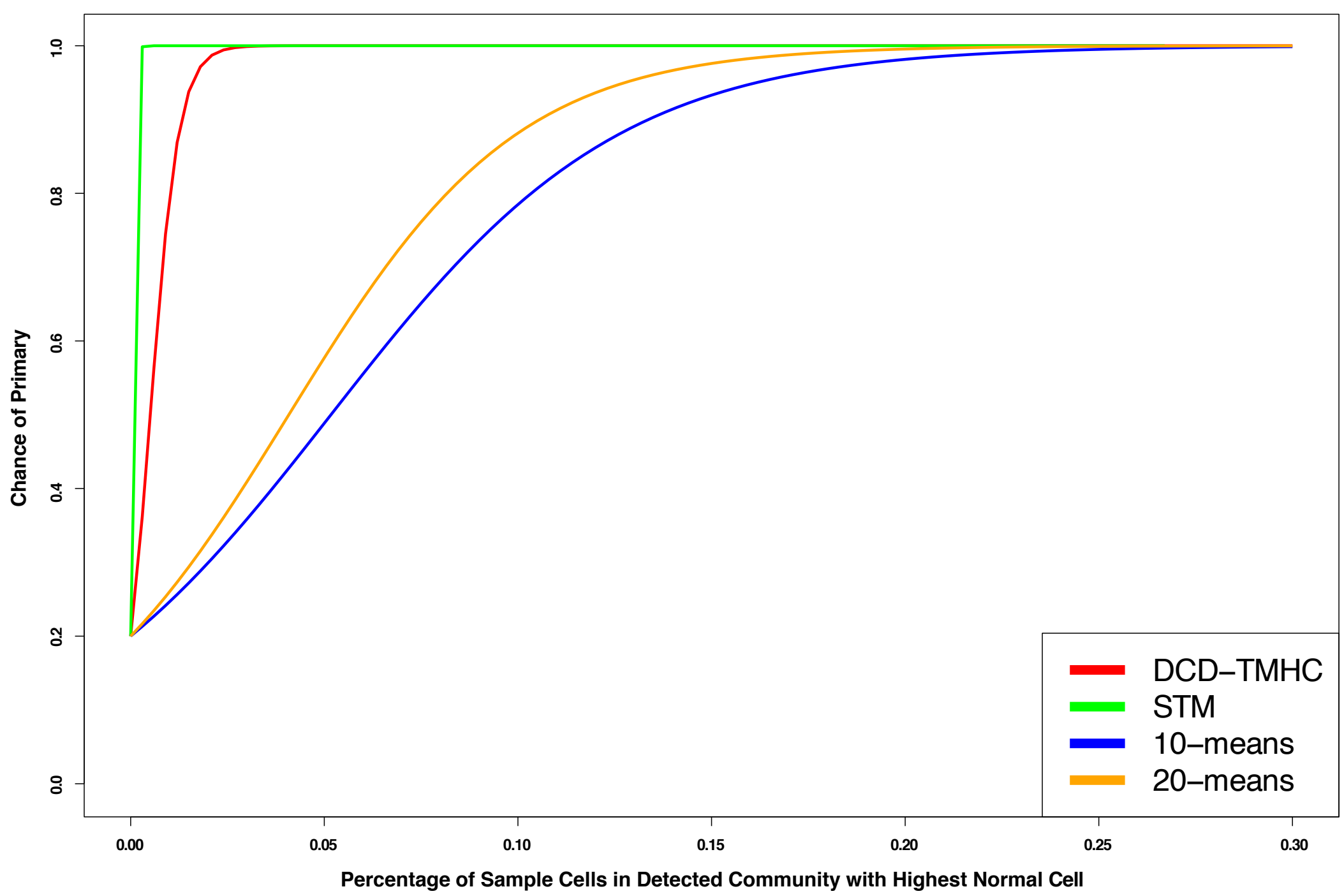

**Remark 3.** ***Logistic Regression Curves in Figure 8.*** From the definition of equations (10)-(12), we see that in Figure 8, four estimated logistic regression curves based on the data in Table 4 are all decreasing functions, and the red curve based on the DCD-TMHC method's detected Community 5 as having the highest percentage of tumor cells is located below all other three curves. The decreasing red curve means that for a sample $S_i$ with larger percentage cells $x_i$ located in Community 5, the patient with cancer tissue sample $S_i$ has less chance of being at primary stage of cancer due to the meaning of $\pi(x)$ and $\hat{\pi}(x)$ given in equations (11)-(12). From $P\{Y = 0 \mid X = x\} = 1 - \pi(x)$, the decreasing red curve also means that for a sample $S_i$ with larger percentage cells $x_i$ located in Community 5, the patient with cancer tissue sample $S_i$ has larger chance of being at metastasis stage of cancer, and our proposed DCD-TMHC method makes such assessment more sharply than all other three methods because the red curve $\hat{\pi}_R(x)$ by DCD-TMHC method being located below all other three curves in Figure 8 means that curve $[1 - \hat{\pi}_R(x)]$ is located above all other three curves $[1 - \hat{\pi}(x)]$; that is for the same large value of $x$, the red curve $[1 - \hat{\pi}_R(x)]$ by DCD-TMHC method predicts largest chance of being at metastasis stage of cancer than all other three methods.

**Remark 4.** ***Logistic Regression Curves in Figure 9 and Figure 10.*** From above Remark 3 on the interpretation of Figure 8, we see clearly that in Figure 9, three estimated logistic regression curves: red $\hat{\pi}_R(x)$, green $\hat{\pi}_G(x)$ and orange $\hat{\pi}_O(x)$ based on data in Table 5 via communities 19, 32 and 12 as having the highest percentages of immune cells detected by methods DCD-TMHC, STM and Elbow $k$-means, respectively, are all increasing functions, while the red curve $\hat{\pi}_R(x)$ by DCD-TMHC method locates significantly above two curves $\hat{\pi}_G(x)$ and $\hat{\pi}_O(x)$. This means that for a sample $S_i$ with large percentage cells $x_i$ located in Community 19 detected by DCD-TMHC method, the patient with cancer tissue sample $S_i$ has large chance of being at primary stage of cancer, and that for the same large value of $x$, the red curve $\hat{\pi}_R(x)$ by DCD-TMHC method predicts largest chance of being at primary stage of cancer than both STM method and Elbow $k$-means method. However, it should be noticed that the blue estimated logistic regression curve $\hat{\pi}_B(x)$ by 10-means method is a slightly decreasing function, which means that 10-means method does not work well here. Thus, our proposed DCD-TMHC method has superior performance than all other methods for detected immune cell related communities. In Figure 10, we see that all four estimated logistic regression curves based on the data in Table 6 are all increasing functions, which are based on the communities as having the highest percentages of normal cells detected by 4 different methods. The curves $\hat{\pi}_R(x)$ and $\hat{\pi}_G(x)$ by DCD-TMHC and STM methods, respectively, are located quite closely, while curves $\hat{\pi}_B(x)$ and $\hat{\pi}_O(x)$ by 10-means and Elbow $k$-means methods, respectively, are also located quite closely. Since curves $\hat{\pi}_R(x)$ and $\hat{\pi}_G(x)$ significantly locate above other two curves, we know that DCD-TMHC and STM methods perform better for detected normal cell related communities.

## 5. Discussion and Conclusion

The spatial transcriptomics (ST) data produced by recent biotechnologies, such as CosMx machine and Xenium machine, contain huge amount of information about cancer tissue samples, which has great potential for improvement of cancer diagnosis and cancer treatment. This article discovers that many existing clustering methods do not work well for community detection of ST data by CosMx, and the commonly used $k$NN compositional data method shows lack of informative neighboring cell patterns for huge CosMx data. Thus, here we propose a novel and more informative *disk compositional data* (DCD) method, which identifies neighboring pattern of each cell via taking into account of the features of ST data produced by recent new technologies.

After initial processing the ST data into disk compositional data matrix, an innovative and interpretable *DCD-TMHC* spatial community detection method is proposed in this paper. Applying various existing methods as well as our DCD-TMHC method, extensive simulation studies and actual analysis of a CosMx breast cancer dataset clearly show that our proposed DCD-TMHC is superior to any other methods.

Based on the spatial communities detected by our proposed DCD-TMHC method for the CosMx breast cancer data, we use logistic regression model to analyze the association and relationship between these identified communities of the CosMx data and cancer research. The results obviously demonstrate that our proposed DCD-TMHC method is clearly interpretable and superior to any other methods, especially in terms of assessment for different stages of cancer.

During upcoming research period, the novel, innovative, informative and interpretable DCD-TMHC method proposed in this article will be helpful and have impact to future cancer research based on ST data for improvement of cancer diagnosis and cancer treatment. In particular, if we can use the new ST technologies to obtain more tissue samples from different types of cancer at different stages, different types of spatial communities can be detected and identified by our DCD-TMHC method for a specific type of cancer at a particular stage, which, by using logistic regression model, appropriate generalized linear models, or other statistical models, can be applied to improve cancer diagnosis and monitor cancer treatment progress.

## Acknowledgments

The author is grateful to the CosMx breast cancer data originally produced by the CosMx machine at *The UNC Lineberger Comprehensive Cancer Center* and correctively processed by *NanoString* company, and would like to thank Dr. Susana Garcia-Recio for processing and providing the data.

## Funding

This work was supported by funds from the NCI Breast SPORE Program P50-CA058223, Breast Cancer Research Foundation BCRF-23-127, and *The UNC Lineberger Triple Negative Breast Cancer Center*. The author was also partially supported by NSF research grant DMS-2113404.